\documentclass[aps,pra,twocolumn,superscriptaddress]{revtex4-1}
\usepackage{graphicx}
\graphicspath{{./figures/}}
\usepackage{amsmath,amsthm,amssymb,physics,color,enumerate,tabularx,makecell,mathrsfs,bbm,stackrel,mathtools,multirow}

\newtheorem{lemma}{Lemma} %to define the environment Lemma with no numbering, use \newtheorem*{}{}

\allowdisplaybreaks %to allow page-breaks inside equations, if needed

\usepackage{soul} %to overstrike words, \st{bla} 

\begin{document}
\title{Continuous-time quantum walks in the presence of a quadratic perturbation}
\author{Alessandro Candeloro}
\email{alessandro.candeloro@unimi.it}
\affiliation{Quantum Technology Lab, Dipartimento di 
Fisica {\em Aldo Pontremoli}, Universit\`{a} degli Studi di Milano, I-20133 Milano, Italy}

\author{Luca Razzoli}
\email{luca.razzoli@unimore.it}
\affiliation{Dipartimento di Scienze Fisiche, Informatiche e 
Matematiche, Universit\`{a} di Modena e Reggio Emilia, I-41125 Modena, 
Italy}

\author{Simone Cavazzoni}
\email{227842@studenti.unimore.it}
\affiliation{Dipartimento di Scienze Fisiche, Informatiche e 
Matematiche, Universit\`{a} di Modena e Reggio Emilia, I-41125 Modena, 
Italy}

\author{Paolo Bordone}
\email{bordone@unimore.it}
\affiliation{Dipartimento di Scienze Fisiche, Informatiche e 
Matematiche, Universit\`{a} di Modena e Reggio Emilia, I-41125 Modena, 
Italy}
\affiliation{Centro S3, CNR-Istituto di Nanoscienze, I-41125 Modena, Italy}

\author{Matteo G. A. Paris}
\email{matteo.paris@fisica.unimi.it}
\affiliation{Quantum Technology Lab, Dipartimento di 
Fisica {\em Aldo Pontremoli}, Universit\`{a} degli Studi di Milano, I-20133 Milano, Italy}
\affiliation{INFN, Sezione di Milano, I-20133 Milano, Italy}

%%%
\begin{abstract}
We address the properties of continuous-time quantum walks with Hamiltonians of the form $\mathcal{H}= L + \lambda L^2$, being $L$ the Laplacian matrix of the underlying graph and being the perturbation $\lambda L^2$ motivated by its potential use to introduce next-nearest-neighbor hopping. We consider cycle, complete, and star graphs because paradigmatic models with low/high connectivity and/or symmetry. First, we investigate the dynamics of an initially localized walker. Then, we devote attention to estimating the perturbation parameter $\lambda$ using only a snapshot of the walker dynamics. Our analysis shows that a walker on a cycle graph is spreading ballistically independently of the perturbation, whereas on complete and star graphs one observes perturbation-dependent revivals and strong localization phenomena. Concerning the estimation of the perturbation, we determine the walker preparations and the simple graphs that maximize the Quantum Fisher Information. We also assess the performance of position measurement, which turns out to be optimal, or nearly optimal, in several situations of interest. Besides fundamental interest, our study may find applications in designing enhanced algorithms on graphs.
\end{abstract}

\date{\today}
\maketitle

\section{Introduction}
\label{sec:intro}
A continuous-time quantum walk (CTQW) describes the dynamics of a quantum particle 
confined to discrete spatial locations, i.e. to the vertices of a graph \cite{farhi1998quantum,childs2002example,wong2016laplacian}. In these systems, the graph Laplacian $L$ (also referred to as the {\em Kirchhoff} matrix 
of the graph) plays the role of the free Hamiltonian, i.e. it 
corresponds to kinetic energy of the particle. Perturbations to ideal CTQW have been investigated earlier \cite{Rao_2011,Caruso_2014,Siloi_2017,Rossi_2017,Cattaneo_2018,Rossi_2018,Herrman_2019,De_Santis_2019,Melnikov_2020}, however with the main focus being about the decoherence effects of stochastic noise, rather than the quantum effects induced by a perturbing Hamiltonian. A notable exception exists, though, given by quantum spatial search, where the perturbation induced by the so-called {\em oracle Hamiltonian} has been largely investigated as a tool to induce localization on a desired site
\cite{childs2004spatial,kendon_2006,Wong_2015,Wong_2016,Philipp_2016,Delvecchio_2020}. As a matter of fact, quantum walks have found several applications ranging from universal quantum computation \cite{childs09} to quantum algorithms \cite{kendon06,ambainis07,farhi08,coppersmith10,Tamascelli_2014,PhysRevE.97.013301}, and to the study of excitation transport on networks \cite{mulken11,alvir16,tamas16}, and  biological systems \cite{mohseni08,hoyer10}. As such, and due to the diversity of the physical platforms on which quantum walks have been implemented \cite{preiss2015strongly,peruzzo2010quantum}, a precise characterization of the quantum-walk Hamiltonian is desired.

\begin{figure}[!b]
\centering	
\includegraphics[width=0.95\columnwidth]{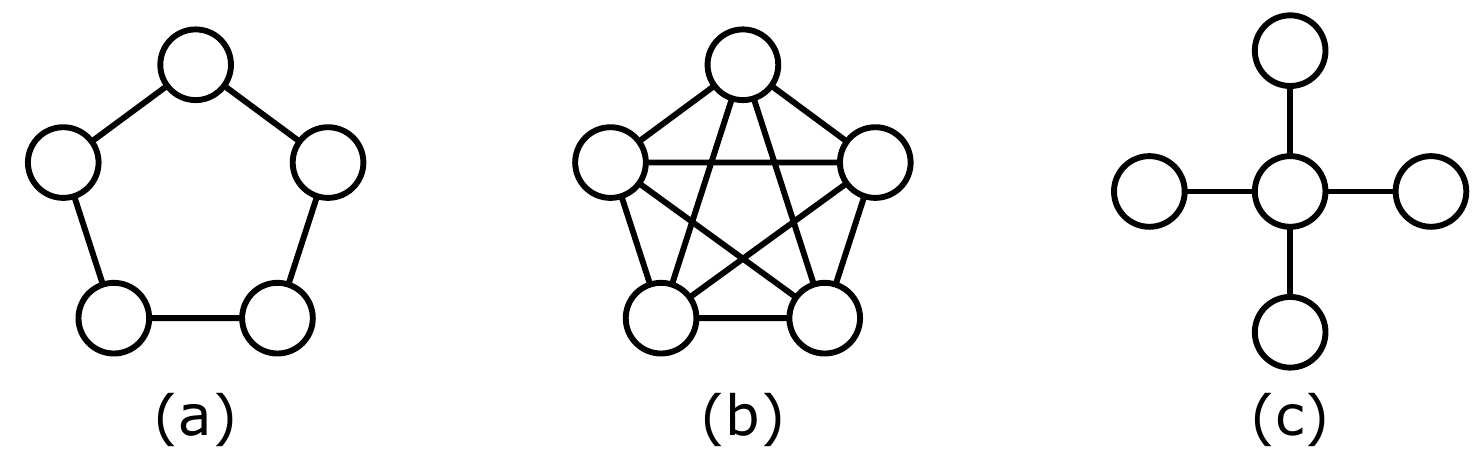}
\caption{The three types of graphs considered in the present work: 
(a) cycle, (b) complete, and (c) star graphs. Examples for $N=5$ vertices.}
\label{fig:graphs}
\end{figure}

In the present work, we investigate the dynamics of an initially localized quantum walker propagating on cycle, complete, and star graphs (see Fig. \ref{fig:graphs}) under perturbed Hamiltonians of the form $\mathcal{H}= L + \lambda L^2$. Characterizing these Hamiltonians amounts to determining the value of the coupling parameter $\lambda$, which quantifies the effects of the quadratic term. For this purpose, we investigate whether, and to which extent, a snapshot of the walker dynamics at a given time suffices to estimate the value of $\lambda$.

Besides the fundamental interest, there are few reasons to address these particular systems. The topologies of these graphs describe paradigmatic situations with low (cycle and star) or high (complete) connectivity, as well as low (star) and high (cycle and complete) symmetry. At the same time, CTQW Hamiltonians with quadratic perturbation of the form $\lambda L^2$ are of interest, e.g., because they represent a physically motivated and convenient way to introduce next-nearest-neighbor hopping in one-dimensional lattices, or intrinsic spin-orbit coupling in two-dimensional ones. Moreover, considering such perturbations is the first step towards the description of dephasing and dechoerence processes, which result from making the parameter $\lambda$ a stochastic process.

To analyze both semi-classical and genuinely quantum features of the dynamics, we employ a set of different quantifiers, including site distribution, mixing, inverse participation ratio, and coherence. In this framework, mixing has been studied for CTQWs on some circulant graphs \cite{ahmadi2003mixing}, e.g. the cycle and the complete graph, and also employed together with the temporal standard deviation to study the dynamics of CTQW on the cycle graph \cite{inui2005evolution}. Moreover, a spectral method has been introduced to investigate CTQW on graphs \cite{jafarizadeh2007investigation,salimi2009continuous}. Coherent transport has been analytically analyzed for CTQW on star graphs \cite{xu2009exact}, showing  the occurrence of perfect revivals and strong localization on the initial node.

The rest of the paper is organized as follows. In Sec. \ref{sec:dynamics} we address the dynamics of an initially localized walker in the different graphs. In Sec. \ref{sec:estimation}, we focus on the estimation of the parameter of the perturbation by evaluating the Quantum Fisher Information (QFI). We consider initially localized states as well as the states maximizing the  QFI, and we compare the QFI to the Fisher Information (FI) of position measurement. Moreover, we determine the simple graphs that allow to obtain the maximum QFI. In Sec. \ref{sec:conclusions}, we summarize and discuss our results and findings. Then, in Appendix \ref{app:an_res_ccsg} we provide further analytical details about the dynamics of the CTQWs over the different graphs. In Appendix \ref{app:q_fi_scg}, we prove the results concerning the (Q)FI.

\section{Dynamics}
\label{sec:dynamics}
A graph is a pair $G=(V, E)$, where $V$ denotes the non-empty set of vertices and 
$E$ the set of edges. In a graph, the kinetic energy term ($\hbar=1$) ${T}=-\nabla^2/2m$ is replaced by ${T}=\gamma L$, where $\gamma \in \mathbb{R}^+$ is the hopping amplitude of the walk and $L=D-A$ the graph Laplacian, with $A$ the adjacency matrix ($A_{jk}=1$ if the vertices $j$ and $k$ are connected, $0$ otherwise) and $D$ the diagonal degree matrix ($D_{jj}=\operatorname{deg}(j)$). The hopping amplitude $\gamma$ plays the role of a time-scaling factor, thus the time dependence of the results is significant when expressed in terms of the dimensionless time $\gamma t$. Please notice that in the following we set $\gamma = \hbar= 1$, and, as a consequence, hereafter time and energy will be dimensionless. We consider finite graphs of order $\abs{V}=N$, i.e. graphs with $N$ vertices which we index from $0$ to $N-1$, and we focus on the dynamics of a walker whose initial state $\ket{\psi(0)}$ is a vertex of the graph, i.e. the walker is initially localized.

We consider the Hamiltonian
\begin{equation}
{\mathcal{H}}={\mathcal{H}}_0+\lambda{\mathcal{H}}_1=L+\lambda L^2\,,
\label{eq:H_ctqwp4}
\end{equation}
where $\lambda$ is a dimensionless perturbation parameter. Because of this choice, the eigenproblem of $\mathcal{H}$ is basically the eigenproblem of $L$. The laplacian eigenvalue $\varepsilon = 0$ is common to all simple graphs, it is not degenerate for connected graphs, like cycle, complete, and star graph, and the corresponding eigenvector is $(1,\ldots,1)/\sqrt{N}$. The time evolution of the system is coherent and ruled by the unitary time-evolution operator
\begin{equation}
\mathcal{U}_\lambda(t)=e^{-i{\mathcal{H}}t}=\sum_{n=0}^{N-1} e^{-i(\varepsilon_n+\lambda\varepsilon_n^2) t} \dyad{e_n}\,,
\label{eq:t_evol_op}
\end{equation}
where the second equality follows from the spectral decomposition of $L$. To study the dynamics of the walker, we consider the following quantities, which basically arise from the density matrix $\rho(t)=\dyad{\psi(t)}$. 

The (\textit{instantaneous}) probability of finding the walker in the vertex $k$ at time $t$ is
\begin{equation}
    P(k,t\vert \lambda)=\abs{\langle k \vert \mathcal{U}_\lambda(t) \vert \psi(0)\rangle}^2\,,
\end{equation}
whereas the \textit{average} probability is
\begin{equation}
     \bar{P}(k\vert \lambda)=\lim_{T\to+\infty}\frac{1}{T}\int_0^T P(k,t\vert \lambda) dt\,.
     \label{eq:avg_prob}
\end{equation}
There are two main notions of mixing in quantum walks \cite{aharonov2001quantum,moore2002quantum,ahmadi2003mixing}. A graph has the \textit{instantaneous exactly uniform mixing} property if there are times when the probability distribution $P(t)$ of the walker is exactly uniform; it has the \textit{average uniform mixing} property if the average probability distribution $\bar{P}$ is uniform.

In addition, we consider the inverse participation ratio (IPR) \cite{thouless1974electrons,kramer1993localization,Rossi_2017}
\begin{equation}
    \mathcal{I}(t) = \sum_{k=0}^{N-1} \langle k \vert \rho (t) \vert k \rangle ^2 = \sum_{k=0}^{N-1} P^2(k,t\vert \lambda)\,,
    \label{eq:ipr_local}
\end{equation}
which allows us to assess the amount of localization in position space of the walker. Indeed, the IPR is bounded from below by $1/N$ (complete delocalization) and from above by $1$ (localization on a single vertex). In this sense, the IPR is an alternative quantity to study the instantaneous exactly uniform mixing. The inverse of the IPR indicates the number of vertices over which the walker is distributed \cite{ingold2002delocalization}.

Finally, to further analyze the quantum features of the dynamics, we consider the quantum coherence. A proper measure is provided by the $l_1$ norm of coherence \cite{baumgratz2014quantifying}
\begin{equation}
\mathcal{C}(t)=\sum_{\substack{j,k=0,\\j\neq k}}^{N-1} \abs{\rho_{j,k}(t)}=\sum_{j,k=0}^{N-1} \abs{\rho_{j,k}(t)}-1\,.
\label{eq:coherence_def}
\end{equation}

Please refer to Appendix \ref{app:an_res_ccsg} for the details about the analytical derivation of the results shown in the following.

\subsection{Cycle graph}
\label{subsec:cyclegraph}
In the cycle graph each vertex is adjacent to $2$ other vertices, so its degree is $2$. Hence, the graph Laplacian is
\begin{equation}
L = 2{I}-\sideset{}{'}\sum_{k=0}^{N-1} \left( \dyad{k-1}{k}+\dyad{k+1}{k}\right)\,.
\label{eq:H0_cycle_matrix}
\end{equation}
The primed summation symbol means that we look at the cycle graph as a path graph provided with periodic boundary conditions, thus the terms $\dyad{-1}{0}$ and $\dyad{N}{N-1}$ are $\dyad{N-1}{0}$ and $\dyad{0}{N-1}$, respectively. The matrix representation of this Laplacian is symmetric and circulant (a special case of Toeplitz matrix), and the related eigenproblem is analytically solved in Ref. \cite{gray2006toeplitz} and reported in Table \ref{tab:eig_pbm_cycle}.   

\begin{table}[tb]
    \centering
    \begin{ruledtabular}
        \begin{tabular}{lcc}
        $\ket{e_n}$ & $\varepsilon_n$ & $\mu_n$\\\hline
        $\ket{e_n}=\frac{1}{\sqrt{N}}\sum\limits_{k=0}^{N-1}e^{-i\frac{2\pi n}{N}k}\ket{k}$ &  $ 2\left[1-\cos\left( \frac{2\pi n}{N} \right) \right]$ & $\ast$\\
        \text{with $n=0,\ldots,N-1$}\\
        \end{tabular}
    \end{ruledtabular}
    \caption{Eigenvectors $\ket{e_n}$ and eigenvalues $\varepsilon_n$ of the graph Laplacian in the cycle graph. ($\ast$) The multiplicity of the eigenvalues depends on the parity of $N$. In particular, the ground state $n=0$ is always unique, whereas the highest energy level is unique for even $N$ and doubly degenerate for odd $N$. Independently of the parity of $N$, the remaining eigenvalues have multiplicity $2$, since $\varepsilon_n = \varepsilon_{N-n}$.}
    \label{tab:eig_pbm_cycle}
\end{table}

The ground state ($n=0$) is unique and equal to
\begin{align}
\varepsilon_{min}&= 0 \label{eq:eval_cyc_min}\,,\\
\ket{e_{min}}&=\frac{1}{\sqrt{N}}\sum_{k=0}^{N-1} \ket{k} \label{eq:evec_cyc_min}\,.
\end{align}
Instead, the highest energy level depends on the parity of $N$ and
\begin{enumerate}[(i)]
\item is unique for even $N$ ($n=N/2$):
\begin{align}
\varepsilon_{max} &=4 \,,\label{eq:eval_cyc_max_even}\\
\ket{e_{max}}&= \frac{1}{\sqrt{N}}\sum_{k=0}^{N-1}(-1)^k\ket{k}\,,\label{eq:evec_cyc_max_even}
\end{align}
\item has degeneracy $2$ for odd $N$ ($n=(N\pm1)/2$):
\begin{align}
\varepsilon_{max} &= 2\left[1+\cos\left (\frac{\pi}{N}\right )\right]\,,\label{eq:eval_cyc_max_odd}\\
\ket{e_{max}}&=
\frac{1}{\sqrt{N}}\sum_{k=0}^{N-1}(-1)^k e^{\pm i \frac{\pi}{N}k}\ket{k}\,,\label{eq:evec_cyc_max_pm}
\end{align}
where the phase factors are all either with the $+$ sign or with the $-$ sign.
\end{enumerate}

Since for odd $N$ the highest energy level is doubly degenerate, we may be interested in finding the corresponding orthonormal eigenstates having real components \footnote{The further reason is that some numerical routines solving the eigenproblem for real symmetric matrices may return orthonormal eigenvectors with real components.}. Therefore we define the following states by linearly combining the two eigenstates in Eq. \eqref{eq:evec_cyc_max_pm} in one case with the plus sign and with the minus sign in the other \footnote{The linear combination leading to Eq. \eqref{eq:evec_cyc_max_sin} introduces also an imaginary unit. However, this is a global phase factor, and, as such, we neglect it.}, respectively:
\begin{align}
\ket{e_{max}^+} &= \sqrt{\frac{2}{N}} \sum_{k=0}^{N-1} (-1)^k \cos \left(\frac{\pi}{N}k\right)\ket{k} \,,\label{eq:evec_cyc_max_cos}\\
\ket{e_{max}^-} &= \sqrt{\frac{2}{N}} \sum_{k=0}^{N-1}(-1)^k \sin \left(\frac{\pi}{N}k\right)\ket{k} \,.\label{eq:evec_cyc_max_sin}
\end{align}

The perturbation involves
\begin{equation}
L^2 =6{I}+ \sideset{}{'}\sum_{k=0}^{N-1} (\dyad{k-2}{k}-4\dyad{k-1}{k}+H.c.)\,,
\label{eq:L2_cyc}
\end{equation}
where the Hermitian conjugate of $\dyad{k-n}{k}$ is a $+n$-vertices hopping term, and, as such, has to be intended as $\dyad{k+n}{k}$. Hence, the perturbed Hamiltonian \eqref{eq:H_ctqwp4} reads as follows:
\begin{align}
{\mathcal{H}}&=(2+6\lambda){I}+\sideset{}{'}\sum_{k=0}^{N-1}\left[ \lambda\dyad{k-2}{k}\right.\nonumber\\
&\quad \left.-(1+4\lambda)\dyad{k-1}{k} +H.c. \right]\,.
\label{eq:totH_cyc}
\end{align}
The perturbation $\lambda L^2$, thus, introduces the next-nearest neighbor hopping, affects the nearest-neighbor one, and also the on-site energies $\propto {I}$.

\begin{figure}[htb]
	\centering	
	\includegraphics[width=0.45\textwidth]{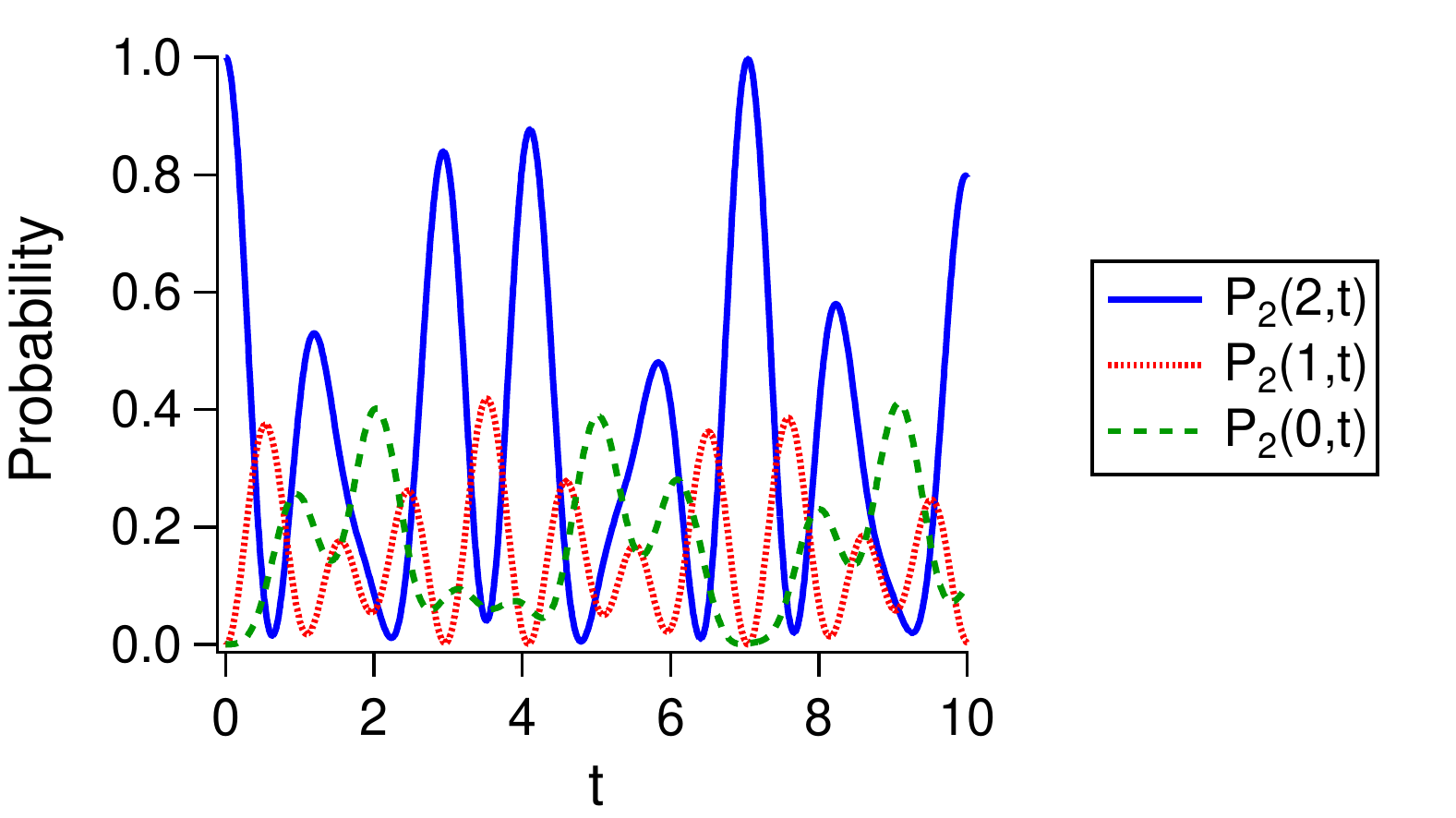}
	\caption{Probability distribution $P_j(k,t\vert\lambda)$ of the walker as a function of time in the cycle graph. The walker is initially localized in the vertex $\ket{j=2}$. The probability distribution is symmetric with respect to the starting vertex, i.e. $P_j(j+k,t\vert\lambda)=P_j(j-k,t\vert\lambda)$. Numerical results suggest that revivals in the starting vertex are most likely not exact. Indeed, to be exact, the periods of the cosine functions entering the definition of the probability \eqref{eq:pjk_cyc_cos} have to be commensurable and such periods strongly depend on the choice of $N$ and $\lambda$. Results for $N=5$ and $\lambda=0.2$.}
	\label{fig:prob_loc_cycle}
\end{figure}

In a cycle graph all the vertices are equivalent, so an initially localized walker will show the same time evolution independently of the starting vertex chosen. We denote the initial state by $\ket{j}$. The probability of finding the walker in the vertex $k$ at time $t$ for a given value of $\lambda$ is (Fig. \ref{fig:prob_loc_cycle})
\begin{align}
    P_j&(k,t\vert\lambda) =\frac{1}{N} + \frac{2}{N^2} \nonumber\\
    & \times \sum_{\substack{n=0, \\ m>n}}^{N-1} \cos\left[(E^\lambda_n - E^\lambda_m)t- \frac{2\pi}{N}(n-m)(j-k)\right]\,,
    \label{eq:pjk_cyc_cos}
\end{align}
which is symmetric with respect to the starting vertex $j$, i.e. $P_j(j+k,t\vert\lambda)=P_j(j-k,t\vert\lambda)$ (proof in Appendix \ref{subapp:an_res_cyc}). The average probability distribution is the same as the one reported in \cite{inui2005evolution}, which is basically our unperturbed CTQW \footnote{The CTQW Hamiltonian in \cite{inui2005evolution} is $\mathcal{H}=A/d$, instead of being the Laplacian. In regular graphs $d$, the degree of the vertex, is the same for all the vertices. The diagonal degree matrix $D$ is thus proportional to the identity, and this introduces an irrelevant phase factor in the time evolution of the quantum state. The time scale of the evolutions under the Hamiltonians $A$ and $A/d$ are clearly different, but the resulting time-averaged probability distribution is the same.}.

The solution of the time-dependent Schr\"{odinger} equation of the unperturbed system ($\lambda=0$) can be expressed in terms of Bessel functions \cite{ahmadi2003mixing}. This allowed to analytically prove the ballistic spreading in a one-dimensional infinite lattice \cite{endo2009ballistic}, i.e. that the variance of the position is $\sigma^2(t) = \langle \hat {x}(t)^2 \rangle - \langle \hat {x}(t) \rangle^2 \propto t^2$.  We expect the same ballistic spreading to characterize the CTQW on a finite cycle at short times, i.e. as long as the walker does not feel the topology of the cycle graph.

We can find a simple expression describing the variance of the position for $\lambda \neq 0$ at short times. The variance is meaningful if we consider sufficiently large $N$, and the assumption $t\ll 1$ ensures that the wavefunction does not reach the vertices $\ket{0}$ and $\ket{N-1}$. Indeed, the position on the graph is the corresponding vertex, but the topology of the cycle graph allows the walker to jump from $\ket{0}$ to $\ket{N-1}$ and \textit{vice versa}. This, in turn, affects the computation of the variance. To ensure the maximum distance from the extreme vertices, we consider a walker initially localized in the central vertex. We assume even $N$, so the starting vertex is $\ket{j=N/2}$. Under these assumptions, we have that
\begin{equation}
\sigma^2 (t) \approx \left[40(\lambda-\lambda_0)^2+\frac{2}{5}\right ]t^2\,,
\label{eq:xVar_cyc}
\end{equation}
with $\lambda_0=-1/5$ (see Appendix \ref{subapp:an_res_cyc}). The spreading of the walker is ballistic in spite of the perturbation. Nevertheless, increasing $\abs{\lambda-\lambda_0}$ makes the walker spread faster by affecting the factor in front of $t^2$. Such factor, indeed, is related to the square of the parameter characterizing the speed of the walker \cite{endo2009ballistic}. The lowest variance is for $\lambda=\lambda_0$, which is the value for which the nearest-neighbor hopping $-(1+4\lambda)$ equals the next-nearest-neighbor one $\lambda$ (see Eq. \eqref{eq:totH_cyc}). Numerical simulations of the CTQW provide evidences that the same behavior in Eq. \eqref{eq:xVar_cyc} characterizes also the CTQW on the cycle with odd $N$ or when the starting vertex is not the central one, again assuming that the wavefunction does not reach the extreme vertices.

\begin{figure}[htb]
	\centering	
	\includegraphics[width=0.45\textwidth]{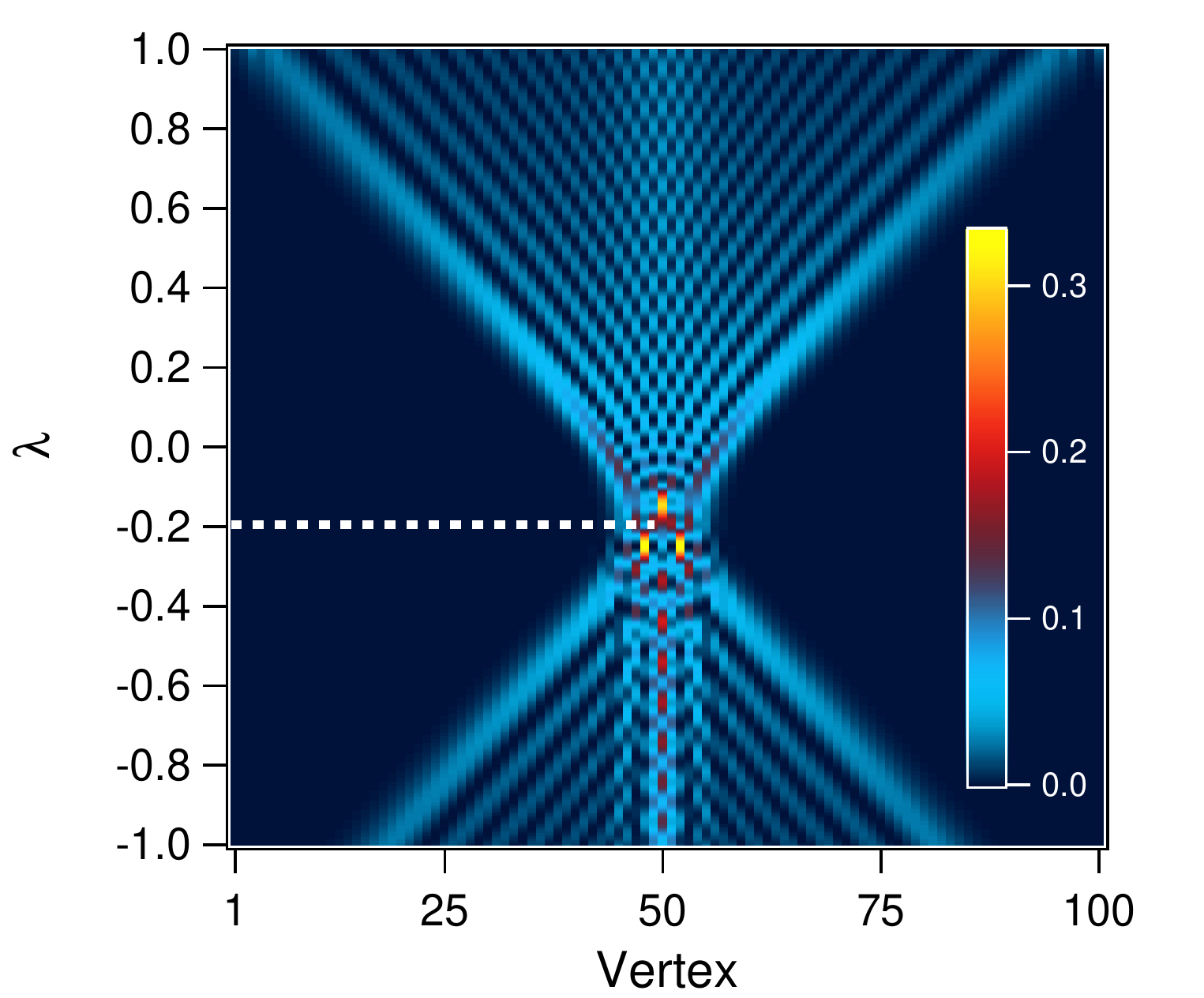}
	\caption{Map of the probability distribution \eqref{eq:pjk_cyc_cos} as a function of the position (vertex) and $\lambda$ at $t=4$. The walker is initially localized in the center of the cycle graph ($N=100$). The horizontal dashed white line highlights $\lambda_0=-1/5$, the value at which the variance of the position is minimum. For clarity, here vertices are indexed from $1$ to $N$.}
	\label{fig:fpd_loc_cycle}
\end{figure}

For completeness, we report in Fig. \ref{fig:fpd_loc_cycle} the numerical results for the probability distribution \eqref{eq:pjk_cyc_cos} at a given time and at varying $\lambda$. The pattern of the probability distribution is not symmetric with respect to $\lambda_0$. Nevertheless, at short times the resulting variance of the position \eqref{eq:xVar_cyc} turns out to be symmetric with respect to $\lambda_0$.

\begin{figure}[htb]
	\centering	
	\includegraphics[width=0.45\textwidth]{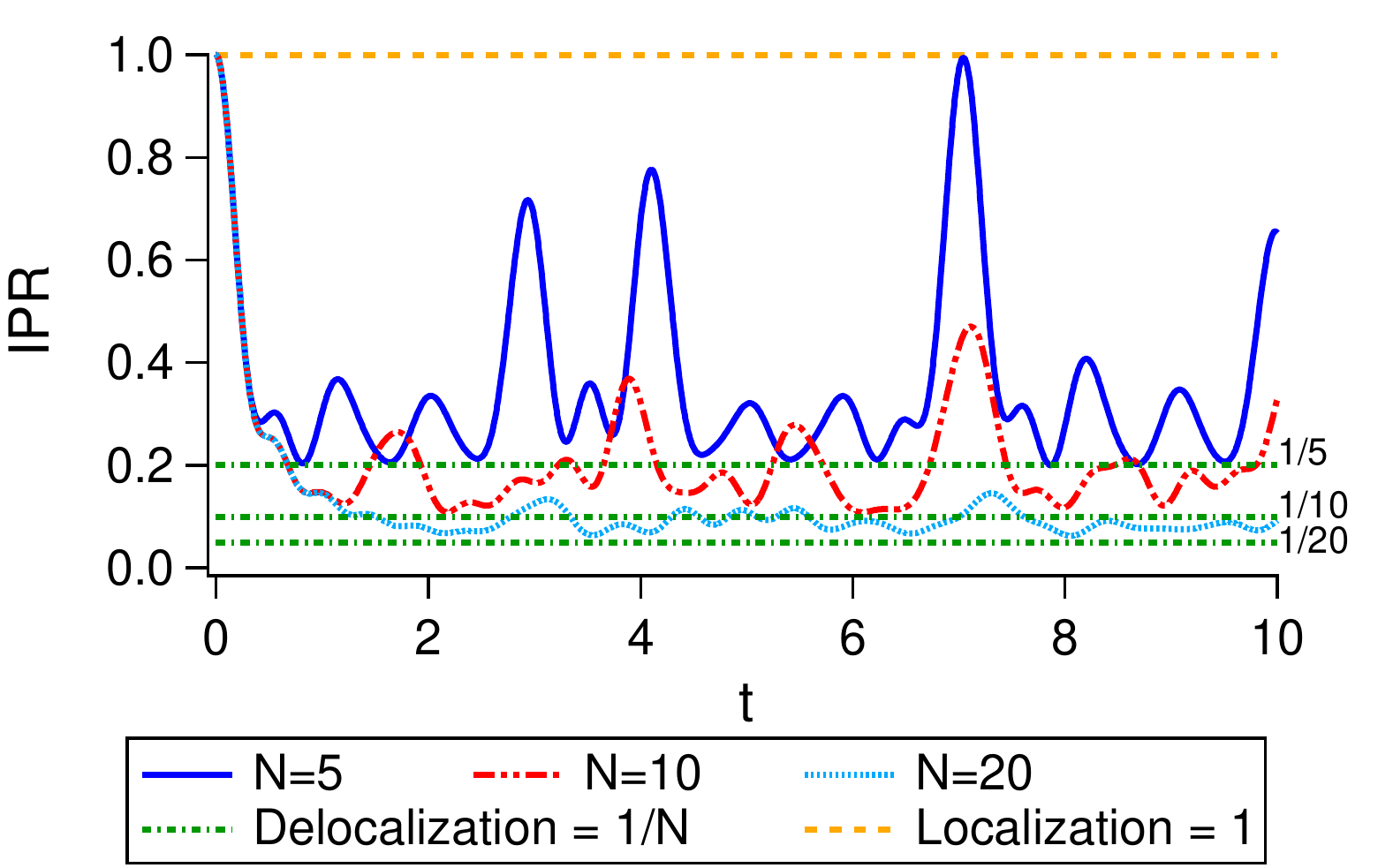}
	\caption{Inverse participation ratio (IPR) for a walker initially localized in the cycle graph. Numerical results suggest that for $t>0$ the IPR reaches neither the lower bound $1/N$ (green dashdotted line), i.e. the delocalization, nor the upper bound $1$ (orange dashed line), i.e. the localization. The fact that the (de)localization is achievable or not is most likely related to the choice of $N$ and $\lambda$. This choice, in turn, might result in the commensurability or incommensurability of the periods of the cosine functions entering the definition of the probability \eqref{eq:pjk_cyc_cos}. For large $N$ the IPR approaches $1/N$, since the probability distribution approaches the uniform one. Results for $\lambda=0.2$.}
	\label{fig:cycle_IPR_multi}
\end{figure}

Next, we numerically evaluate the IPR \eqref{eq:ipr_local} for the probability distribution in Eq. \eqref{eq:pjk_cyc_cos}, and the results are shown in Fig. \ref{fig:cycle_IPR_multi}. As expected from the previous results about the probability distribution (see also Fig. \ref{fig:prob_loc_cycle}), the IPR does not show a clear periodicity, it strongly fluctuates, and there are instants of time when it gets closer to $1$, meaning that the walker is more localized. The numerical results also suggest that the instantaneous exactly uniform mixing is achievable for $N\leq 4$, while there is no exact delocalization for $N>4$, as already conjectured \cite{ahmadi2003mixing}. However, for large $N$ the probability distribution \eqref{eq:pjk_cyc_cos} approaches the uniform one, and so the IPR approaches $1/N$.

\begin{figure}[htb]
	\centering	
	\includegraphics[width=0.45\textwidth]{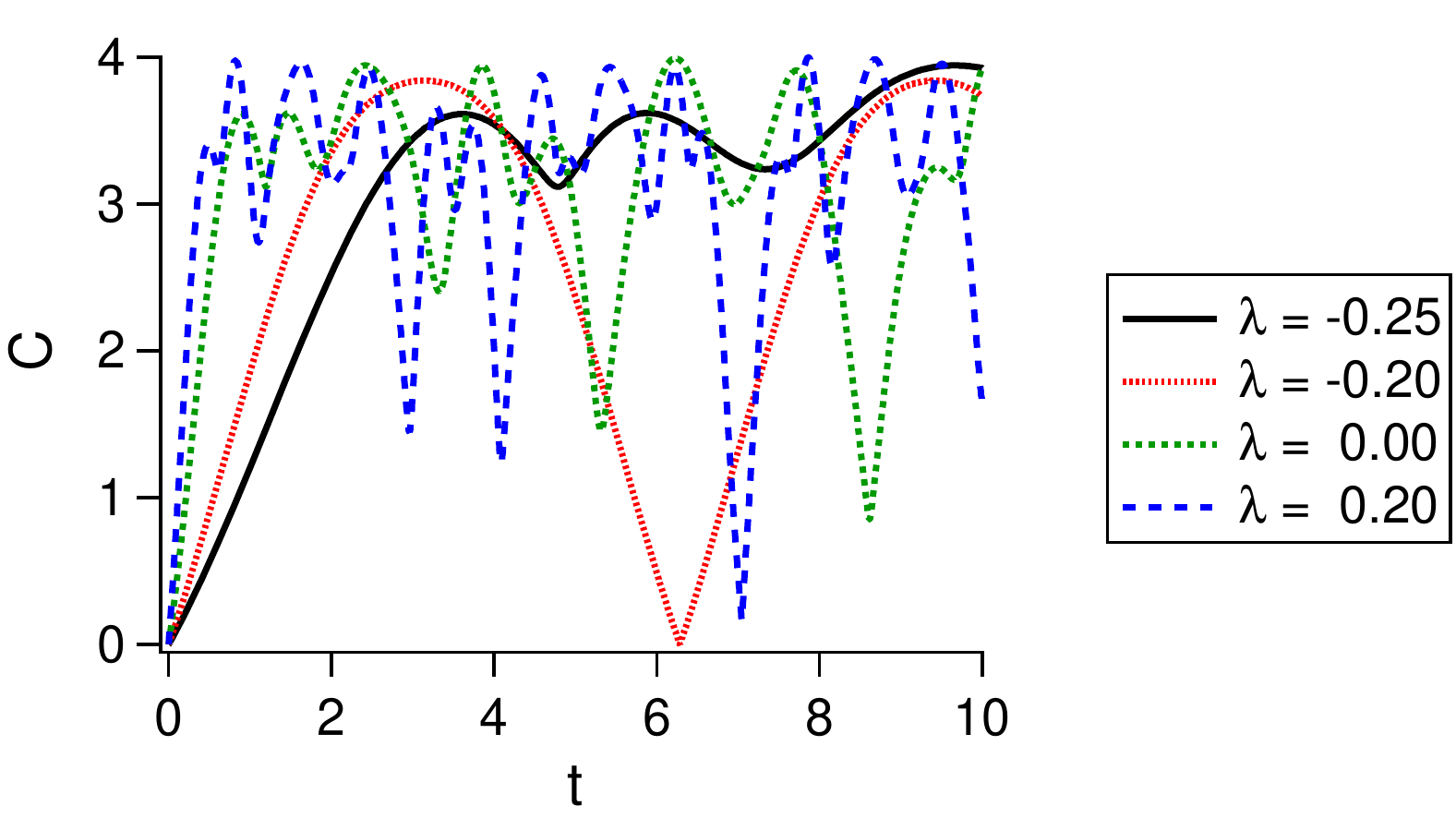}
	\caption{Coherence for a walker initially localized in the cycle graph with $N=5$. For $t\ll 1$ the minimum is for $\lambda =-1/4$, as expected from the linear approximation in Eq. \eqref{eq:coher_smallt_cycle}.}
	\label{fig:coher_cyc_loc}
\end{figure}

Finally, we focus on the time dependence of the coherence \eqref{eq:coherence_def} for an initially localized walker. The exact numerical results are shown in Fig. \ref{fig:coher_cyc_loc}. Under the assumption $t \ll 1$, we can find a simple expression. We Taylor expand the time-evolution operator up to the first order, so the density matrix is approximated as ${\rho}(t) = {\rho}(0)-it\left[\mathcal{H},\rho(0)\right]+\mathcal{O}(t^2)$. Then, being the Hamiltonian \eqref{eq:totH_cyc}, the behavior characterizing the earlier steps of the time evolution of the coherence is
\begin{equation}
    \mathcal{C}(t,\lambda) \approx 4(\abs{\lambda}+\abs{1+4\lambda} )t\,,
    \label{eq:coher_smallt_cycle}
\end{equation}
consistently with the results shown in Fig. \ref{fig:coher_cyc_loc}. Hence, at short times the coherence is minimum for $\lambda = -1/4$. For such value the nearest-neighbor hopping $-(1+4\lambda)$ is null, while the next-nearest-neighbor hopping $\lambda$ is nonzero (see Eq. \eqref{eq:totH_cyc}).

\subsection{Complete graph}
\label{subsec:completegraph}
In the complete graph each vertex is adjacent to all the others, so its degree is $N-1$. Hence, the graph Laplacian is
\begin{equation}
    L = (N-1){I}-\sum_{\substack{j,k=0,\\j\neq k}}^{N-1} \dyad{j}{k}\,,
    \label{eq:H0_complete_matrix}
\end{equation}
and has the following property
\begin{equation}
L^n =N^{n-1}L\,.
\label{eq:Ln_proptoL}
\end{equation}

\begin{table}[tb]
    \centering
    \begin{ruledtabular}
        \begin{tabular}{clcc}
        $n$ & $\ket{e_n}$ & $\varepsilon_n$ & $\mu_n$\\\hline
        $0$ & $\ket{e_0}=\frac{1}{\sqrt{N}}\sum\limits_{k=0}^{N-1}\ket{k}$ &  $0$ & $1$ \\
        $1$ & $\ket{e_1^l}=\frac{1}{\sqrt{l(l+1)}}\left(\sum\limits_{k=0}^{l-1}\ket{k}-l\ket{l}\right)$ &  $N$ & $N-1$\\
        &\text{with $l=1,\ldots,N-1$}\\
        \end{tabular}
    \end{ruledtabular}
    \caption{Eigenvectors $\ket{e_n}$, eigenvalues $\varepsilon_n$ with multiplicity $\mu_n$ of the graph Laplacian in the complete graph.}
    \label{tab:eig_pbm_complete}
\end{table}

The eigenproblem related to Eq. \eqref{eq:H0_complete_matrix} is solved in Table \ref{tab:eig_pbm_complete}. The graph Laplacian has two energy levels:
\begin{enumerate}[(i)]
    \item the level $\varepsilon_0 = 0$, having eigenstate $\ket{e_0}$;
    \item the $(N-1)$-degenerate level $\varepsilon_1 = N$, having orthonormal eigenstates $\ket{e_1^l}$, with $l=1,\ldots,N-1$.
\end{enumerate}

The perturbed Hamiltonian is therefore
\begin{equation}
    \mathcal{H}=(1+N\lambda)L\,,
    \label{eq:totH_comp}
\end{equation}
i.e. it is basically the CTQW Hamiltonian of the complete graph multiplied by a constant which linearly depends on $\lambda$. We observe that the value $\lambda^\ast=-1/N$ makes the Hamiltonian null, and so it makes this case trivial. The perturbation affects the energy scale of the unperturbed system, and so its time scale. Therefore, we can directly compare the next results with the well-known ones concerning the unperturbed system \cite{ahmadi2003mixing}.

The time evolution operator \eqref{eq:t_evol_op} is
\begin{equation}
e^{-i \mathcal{H} t}=I+\frac{1}{N}\left [e^{-i2\omega_N(\lambda)t}-1\right ]L\,.
\end{equation}
where we have Taylor expanded the exponential, used Eq. \eqref{eq:Ln_proptoL}, and defined the angular frequency
\begin{equation}
    \omega_N(\lambda)=\frac{N}{2}(1+\lambda N)\,,
    \label{eq:ang_freq}
\end{equation}
which depends on $\lambda$.  For large $N$, the time evolution basically results in adding a phase to the initial state, since $\lim_{N\to+\infty} \mathcal{U}_\lambda(t)=\exp\left[-i2\omega_N(\lambda)t\right] I$.  

\begin{figure}[htb]
	\centering	
	\includegraphics[width=0.45\textwidth]{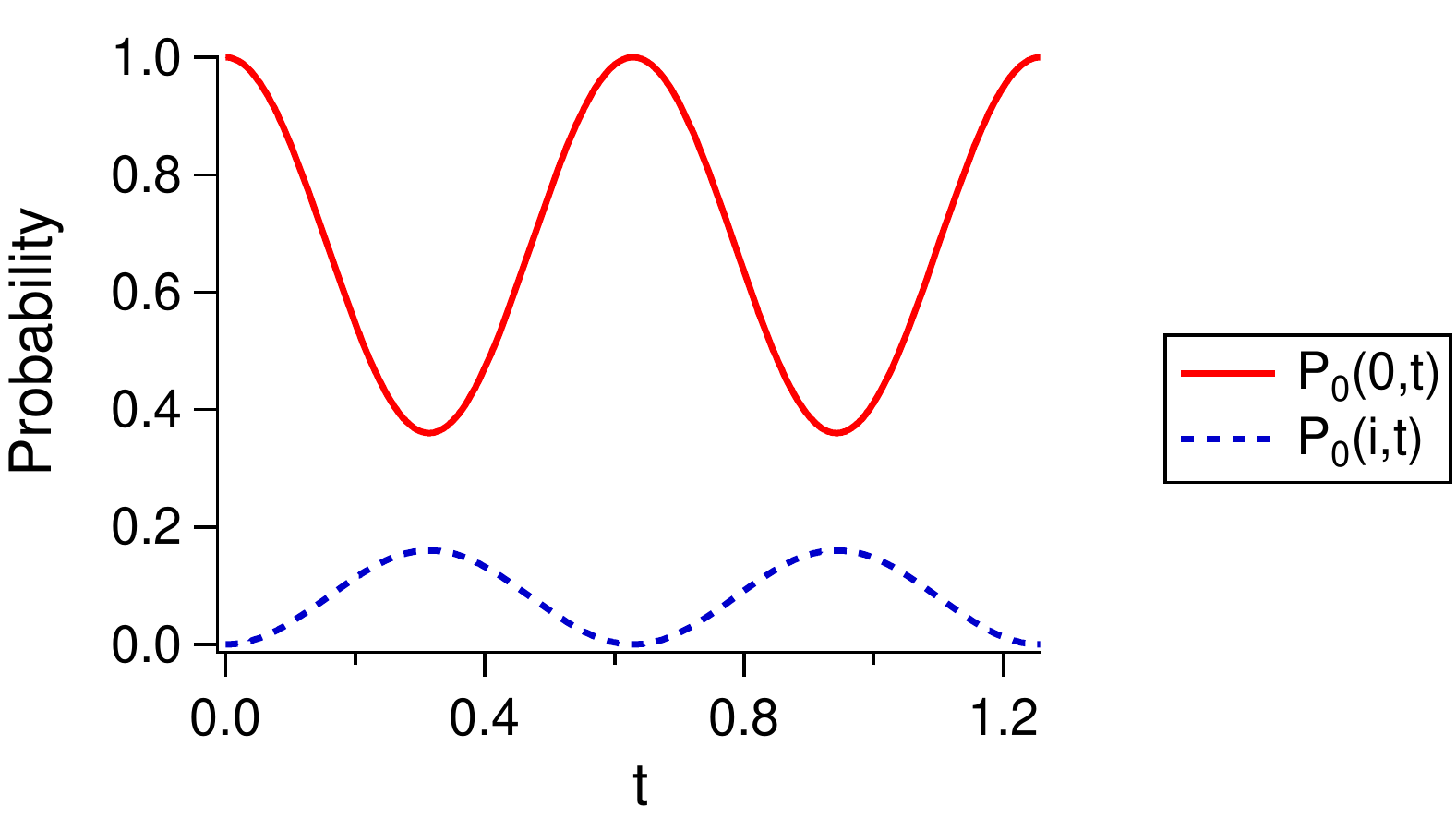}
	\caption{Probability of finding the walker in the starting vertex $P_0(0,t\vert\lambda)$ (red solid line) or in any other vertex $P_0(i,t\vert\lambda)$ (blue dashed line) as a function of time in the complete graph. The walker is initially localized in the vertex $\ket{0}$. Results for $N=5$ and $\lambda=0.2$.}
	\label{fig:prob_loc_complete}
\end{figure}

In a complete graph all the vertices are equivalent, so an initially localized walker will show the same time evolution independently of the starting vertex chosen. We denote the initial state by $\ket{0}$. The probabilities of finding the walker in $\ket{0}$ or elsewhere, $\ket{1\leq i \leq N-1}$, at time $t$ for a given value of $\lambda$ are periodic (Fig. \ref{fig:prob_loc_complete})
\begin{align}
    P_0(0,t\vert\lambda) & = 1 - \frac{4(N-1)}{N^2} \sin^2\left(\omega_N(\lambda)t\right), \label{eq:p0_cg}\\
    P_0(i,t\vert \lambda) & = \frac{4}{N^2}\sin^2\left(\omega_N(\lambda)t\right)\,. \label{eq:pi_cg}
\end{align}
Hence, the walker comes back periodically to the starting vertex and can be found in it with certainty. This occurs for $t_k= 2 k \pi /(N+\lambda N^2)$, with $k\in \mathbb{N}$. Increasing the order of the graph makes the angular frequency higher, and $\lim_{N\to+\infty}P_0(0,t\vert\lambda) = 1$, while $\lim_{N\to+\infty}P_0(i,t\vert\lambda) = 0$. The perturbation only affects the periodicity of the probabilities. The probability distribution is symmetric with respect to $\lambda^\ast$, since $\omega_N(\lambda^\ast \pm\lambda)=\pm \lambda N^2 / 2$ and $\sin^2\left( \lambda N^2 / 2 \right)=\sin^2\left(-\lambda N^2 / 2 \right)$. As expected, for $\lambda^\ast$ the walker remains in the starting vertex all the time, since $\omega_N(\lambda^\ast)=0$ and so $P_0(0,t\vert\lambda^\ast) = 1\,\forall\,t$. The average probability distribution is the same as the one reported in \cite{ahmadi2003mixing}, which is basically our unperturbed CTQW \footnote{The CTQW Hamiltonian in \cite{ahmadi2003mixing}  is $\mathcal{H}=A/d$. See also footnote \cite{Note3}.}.

\begin{figure}[htb]
	\centering	
	\includegraphics[width=0.45\textwidth]{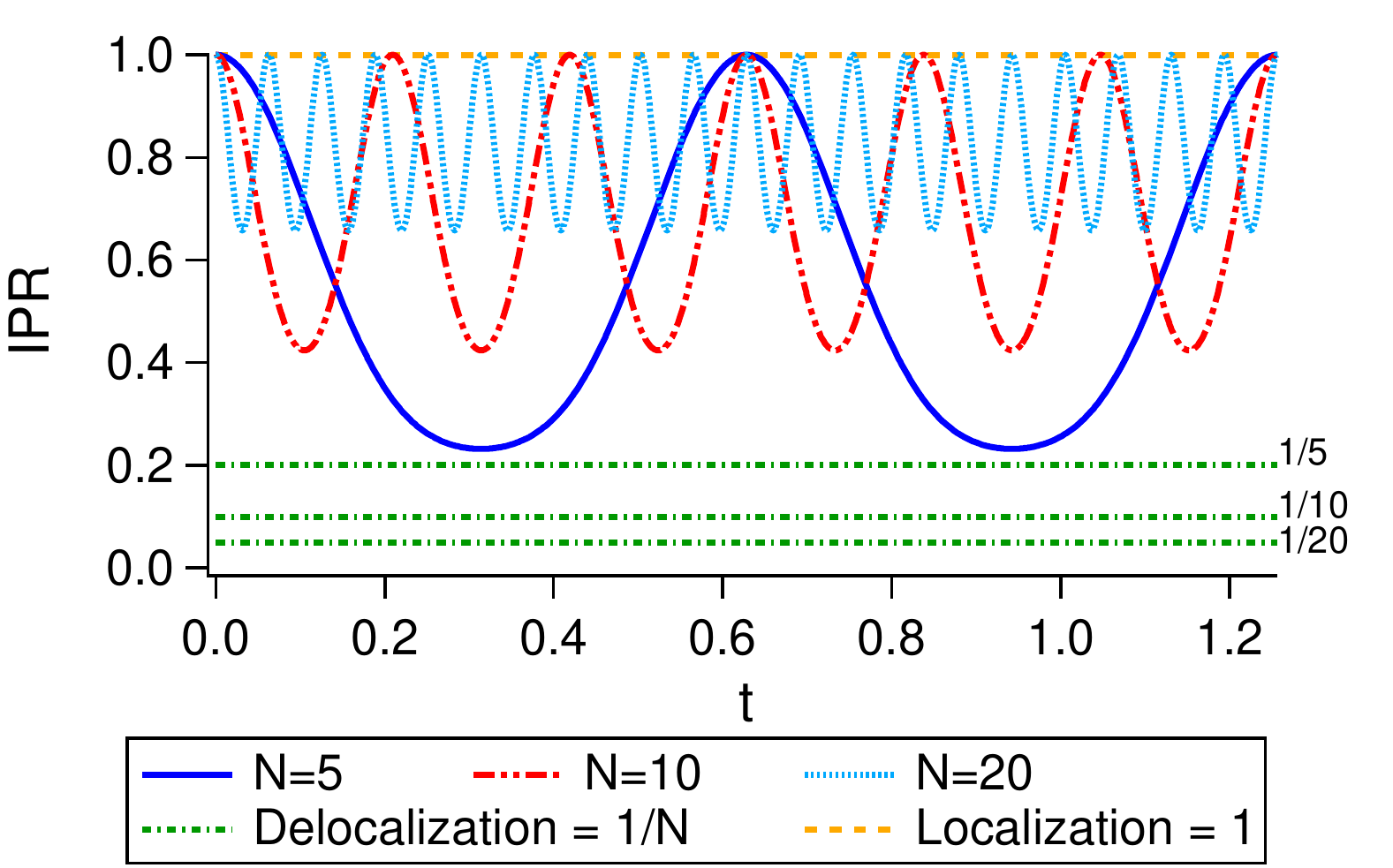}
	\caption{Inverse participation ratio (IPR) for a walker initially localized in the complete graph. The IPR periodically reaches the upper bound $1$ (orange dashed line), i.e. the localization, but for $N>4$ does not reach the value $1/N$ (green dashdotted line), i.e. the delocalization. The lower bound of the IPR is defined in Eq. \eqref{eq:low_bound_cg_IPR}. For $N\to+\infty$ the IPR approaches 1, since the probability of finding the walker in the starting vertex approaches $1$ (see Eqs. \eqref{eq:p0_cg}--\eqref{eq:pi_cg}). Results for $\lambda=0.2$.}
	\label{fig:cg_IPR_multi}
\end{figure}

Next, the IPR \eqref{eq:ipr_local} for the probability distribution in Eqs. \eqref{eq:p0_cg}--\eqref{eq:pi_cg} reads as
\begin{align}
    \mathcal{I}(t) = & 1 - \frac{8(N-1)}{N^2}\sin^2(\omega_N(\lambda)t)\nonumber \\
    & + \frac{16(N-1)}{N^3}\sin^4(\omega_N(\lambda)t)\,.
\end{align}
The IPR has the same properties of the probability distribution: it is periodic, reaches the upper bound $1$ (localization of the walker) for $t_k$ such that $P_0(0,t_k\vert \lambda)=1$,  and $\lim_{N \to + \infty}\mathcal{I}=1$, since for large $N$ the walker tends to be localized in the starting vertex (Fig. \ref{fig:cg_IPR_multi}). The lower bound $\mathcal{I}_{m}:=\min_t \mathcal{I}$ actually depends on $N$:
\begin{equation}
    \mathcal{I}_{m} =\mathcal{I}(t_l)= \begin{dcases}
    \frac{1}{N} & \text{for $N\leq 4$}\,,\\
    1-\frac{8}{N}+\frac{24}{N^2}-\frac{16}{N^3} & \text{for $N>4$}\,,
    \end{dcases}
    \label{eq:low_bound_cg_IPR}
\end{equation}
where
%\begin{equation}
%    t_l \begin{dcases}
%    \text{s.t. }\sin^2(\omega_N(\lambda)t_l) = \frac{N}{4} & \text{for $N\leq 4$}\,,\\
%    =\frac{2\pi(1/2 + l)}{N+\lambda N^2} & \text{for $N>4$}\,,
%    \end{dcases}
%\end{equation}
\begin{equation}
    t_l =\begin{dcases}
    \frac{2 [\pm\arcsin{(\sqrt{N}/2)}+\pi l]}{N+\lambda N^2} & \text{for $N\leq 4$}\,,\\
    \frac{2\pi(1/2 + l)}{N+\lambda N^2} & \text{for $N>4$}\,,
    \end{dcases}
\end{equation}
with $l\in \mathbb{N}$. Please notice that the two definitions of $\mathcal{I}_m$ match in $N=4$. For $N\leq 4$ there are instants of time when the walker is delocalized ($\mathcal{I}_m=1/N$) and there is instantaneous exactly uniform mixing. Instead, for $N>4$ the walker is never delocalized, since $\mathcal{I}_m>1/N$. 

Finally, we focus on the time dependence of the coherence, which we derive in Appendix \ref{subapp:an_res_cg} and is shown in Fig. \ref{fig:coher_complete_loc}. The modulus of the off-diagonal elements of the density matrix can be expressed in terms of the square root of probabilities (see Appendix \ref{app:an_res_ccsg}), thus the coherence is periodic and it is symmetric with respect to $\lambda^\ast$, as well as the probability distribution.  As expected, the dependence on the perturbation is encoded only in the angular frequency $\omega_n(\lambda)$, and the coherence is identically null, thus minimum, for $\lambda^\ast$. For $\lambda \neq \lambda^\ast$, the coherence periodically reaches the following extrema
\begin{align}
    \max \mathcal{C}&=\frac{8(N-1)(N-2)}{N^2} \quad \text{for  } t_k = \frac{(2k+1)\pi}{N+\lambda N^2}\,,\\
    \min \mathcal{C}&=0\quad \text{for  } t_k = \frac{2k\pi}{N+\lambda N^2}\,,
\end{align}
with $k\in\mathbb{N}$, and assuming $N\geq 2$.

\begin{figure}[htb]
	\centering	
	\includegraphics[width=0.45\textwidth]{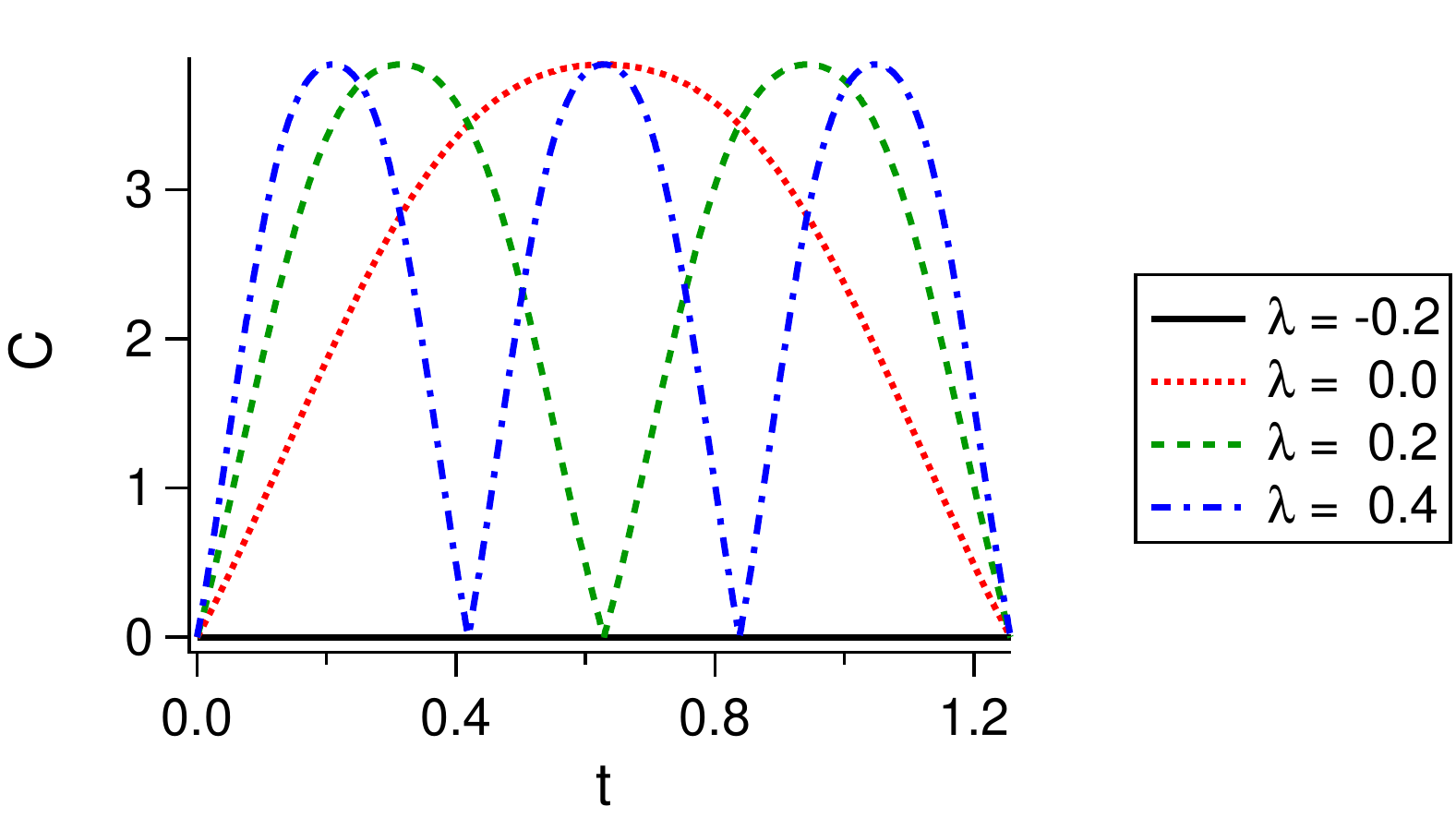}
	\caption{Coherence for a walker initially localized in the complete graph with $N=5$. The coherence is null, thus minimum, for $\lambda^\ast =-1/N$, and it is symmetric with respect to $\lambda^\star$, so only the data for $\lambda\geq\lambda^\ast$ are shown.}
	\label{fig:coher_complete_loc}
\end{figure}

\subsection{Star graph}
\label{subsec:stargraph}
In the star graph, the central vertex is adjacent to all the others, so its degree is $N-1$. On the other hand, the other vertices are only connected to the central one, thus their degree is $1$. Hence, the graph Laplacian is
\begin{equation}
L=I+(N-2)\dyad{0}-\sum_{k=1}^{N-1}(\dyad{k}{0}+\dyad{0}{k})\,,
\label{eq:H0_star_matrix}
\end{equation}
where $\ket{0}$ denotes the central vertex.
 
 \begin{table}[tb]
    \centering
    \begin{ruledtabular}
        \begin{tabular}{clcc}
        $n$ & $\ket{e_n}$ & $\varepsilon_n$ & $\mu_n$\\\hline
        $0$ & $\ket{e_0}=\frac{1}{\sqrt{N}}\sum\limits_{k=0}^{N-1}\ket{k}$ &  $0$ & $1$ \\
        $1$ & $\ket{e_1^l}=\frac{1}{\sqrt{l(l+1)}}\left(\sum\limits_{k=1}^{l}\ket{k}-l\ket{l+1}\right)$ &  $1$ & $N-2$\\
        &\text{with $l=1,\ldots,N-2$}\\
        $2$ & $\ket{e_2}=\frac{1}{\sqrt{N(N-1)}}\left[(N-1)\ket{0}-\sum\limits_{k=1}^{N-1}\ket{k}\right]$ &  $N$ & $1$\\
        \end{tabular}
    \end{ruledtabular}
    \caption{Eigenvectors $\ket{e_n}$, eigenvalues $\varepsilon_n$ with multiplicity $\mu_n$ of the graph Laplacian in the star graph.}
    \label{tab:eig_pbm_star}
\end{table}

The eigenproblem related to Eq. \eqref{eq:H0_star_matrix} is solved in Table \ref{tab:eig_pbm_star}. The graph Laplacian has three energy levels:
\begin{enumerate}[(i)]
    \item the level $\varepsilon_0 = 0$, having eigenstate $\ket{e_0}$;%, is common to all simple graphs and it is not degenerate since the star graph is connected;
    \item the $(N-2)$-degenerate level $\varepsilon_1 = 1$, having orthonormal eigenstates $\ket{e_1^l}$, with $l=1,\ldots,N-2$;
    \item the level $\varepsilon_1 = N$, having eigenstate $\ket{e_2}$.
\end{enumerate}
 
The perturbation involves
\begin{align}
L^2&=2I+(N^2-N-2)\dyad{0}\nonumber\\
&-N\sum_{k=1}^{N-1}(\dyad{k}{0}+\dyad{0}{k})+\sum_{\substack{j,k=1,\\j\neq k}}^{N-1}\dyad{j}{k}\,,
\label{eq:L2_star}
\end{align}
so the perturbed Hamiltonian \eqref{eq:H_ctqwp4} reads as follows:
\begin{align}
{\mathcal{H}}&=(1+2\lambda)I+[N-2+\lambda(N^2-N-2)]\dyad{0}\nonumber\\
&-(1+\lambda N)\sum_{k=1}^{N-1}(\dyad{k}{0}+\dyad{0}{k})+\lambda\sum_{\substack{j,k=1,\\j\neq k}}^{N-1}\dyad{j}{k}\,.
\label{eq:totH_star}
\end{align}
The perturbation $\lambda L^2$, thus, introduces the hopping among all the outer vertices (next-nearest neighbors), affects the hopping to and from the central vertex, i.e. the nearest-neighbor hopping, and also the on-site energies $\propto {I}$.

For an initially localized state, there are two different time evolutions. If at $t=0$ the walker is in the central vertex  $\vert 0 \rangle$, then the time evolution is equal to the corresponding one in the complete graph of the same size. Therefore, also the resulting probability distribution, the IPR, and the coherence are equal between star and complete graph. Instead, if at $t=0$ the walker is localized in any of the outer vertices, then we have a different time evolution. All the outer vertices $\ket{1 \leq i \leq N -1}$ are equivalent and differ from the central vertex $\ket{0}$, so, if we keep the central vertex as $\ket{0}$, we can always relabel the outer vertices in such a way that the starting vertex is denoted by $\ket{1}$.

\begin{figure}[htb]
	\centering	
	\includegraphics[width=0.45\textwidth]{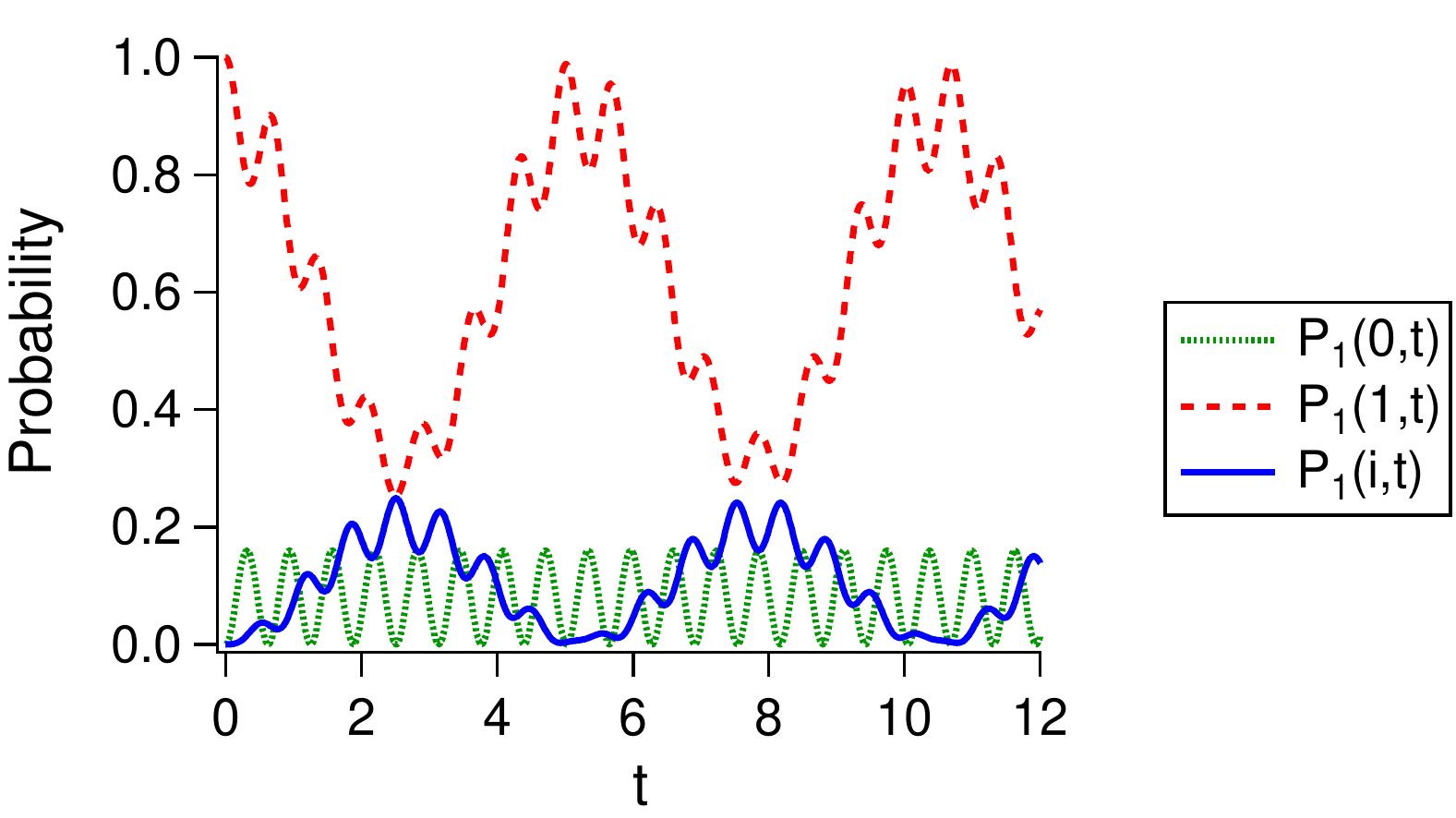}
	\caption{Probability of finding the walker in the central vertex $P_1(0,t\vert\lambda)$ (green dotted line), in the starting vertex $P_1(1,t\vert\lambda)$ (red dashed line) or in any other vertex $P_1(i,t\vert\lambda)$ (blue solid line) as a function of time in the star graph. The walker is initially localized in the vertex $\ket{1}$. Results for $N=5$ and $\lambda=0.2$.}
	\label{fig:prob_loc_star}
\end{figure}

The probabilities of finding the walker in the central vertex $\ket{0}$, in the starting vertex $\ket{1}$ or in any other outer vertex $\ket{2\leq i\leq N-1}$ at time $t$ for a given value of $\lambda$ are respectively (Fig. \ref{fig:prob_loc_star})
\begin{align}
    P_1(0,t\vert \lambda) & = \frac{4}{N^2}\sin^2(\omega_N(\lambda)t)\,, \label{eq:p_10_sg}\\
    P_1(1,t\vert \lambda) & = 1 - \frac{4}{N(N-1)} \Big[(N-2)\sin^2(\omega_1(\lambda)t) \nonumber\\
    & + \frac{N-2}{N-1} \sin^2[(\omega_{N}(\lambda )-\omega_1(\lambda))t] \nonumber \\
    & + \frac{1}{N} \sin^2(\omega_N(\lambda)t)\Big]\,,\label{eq:p_11_sg}\\
    P_1(i,t\vert \lambda) & = \frac{4}{N(N-1)} \Big[ \sin^2(\omega_1(\lambda)t) \nonumber \\
    & + \frac{1}{N-1} \sin^2[(\omega_{N}(\lambda )-\omega_1(\lambda))t] \nonumber \\
    & -\frac{1}{N} \sin^2(\omega_N(\lambda)t)\Big]\label{eq:p_1i_sg}\,,
\end{align}
where the angular frequency is defined in Eq. \eqref{eq:ang_freq}. In particular, $P_1(0,t\vert \lambda)$ is periodic with period $T_N:=\pi / \omega_N(\lambda)$, it is symmetric with respect to $\lambda^\ast=-1/N$, and $P_1(0,t\vert \lambda^\ast)=0$, which means that the walker lives only in the outer vertices of the star graph. Indeed, $\lambda^\ast$ makes the hopping terms to and from the central vertex $\ket{0}$ null (see Eq. \eqref{eq:totH_star}). Instead, $P_1(1,t\vert \lambda)$ and $P_1(i,t\vert \lambda)$ are periodic if and only if the periods $T_1$, $T_N$, and $\pi/(\omega_N(\lambda)-\omega_1(\lambda))$ of the summands are commensurable.
When this happens, then the overall probability distribution is periodic. This happens also for the particular values $\lambda=-1,-1/N,-1/(N+1)$, which make null $\omega_1$, $\omega_N$, and $\omega_N-\omega_1$, respectively. Indeed, when $\omega_1$ ($\omega_N$) is null, the probabilities \eqref{eq:p_10_sg}--\eqref{eq:p_1i_sg} only involve sine functions with $\omega_N$ ($\omega_1$). When $\omega_N-\omega_1=0$, i.e. $\omega_N=\omega_1$, all the sine functions have the same angular frequency. We address in details the periodicity of the probability distribution in Appendix \ref{subapp:an_res_sg}. For $P_1(1,t\vert \lambda)$ and $P_1(i,t\vert \lambda)$ results suggest that there is no symmetry with respect to $\lambda$. Increasing the order of the graph makes the angular frequency higher, and $\lim_{N\to+\infty}P_1(1,t\vert\lambda) = 1$, while $\lim_{N\to+\infty}P_1(0,t\vert\lambda)=\lim_{N\to+\infty}P_1(i,t\vert\lambda) = 0$. Again, the perturbation affects the probabilities only through the angular frequency. The average probability distribution is the same as the one reported in \cite{xu2009exact}, which is exactly our unperturbed CTQW.
%%%

\begin{figure}[htb]
	\centering	
	\includegraphics[width=0.45\textwidth]{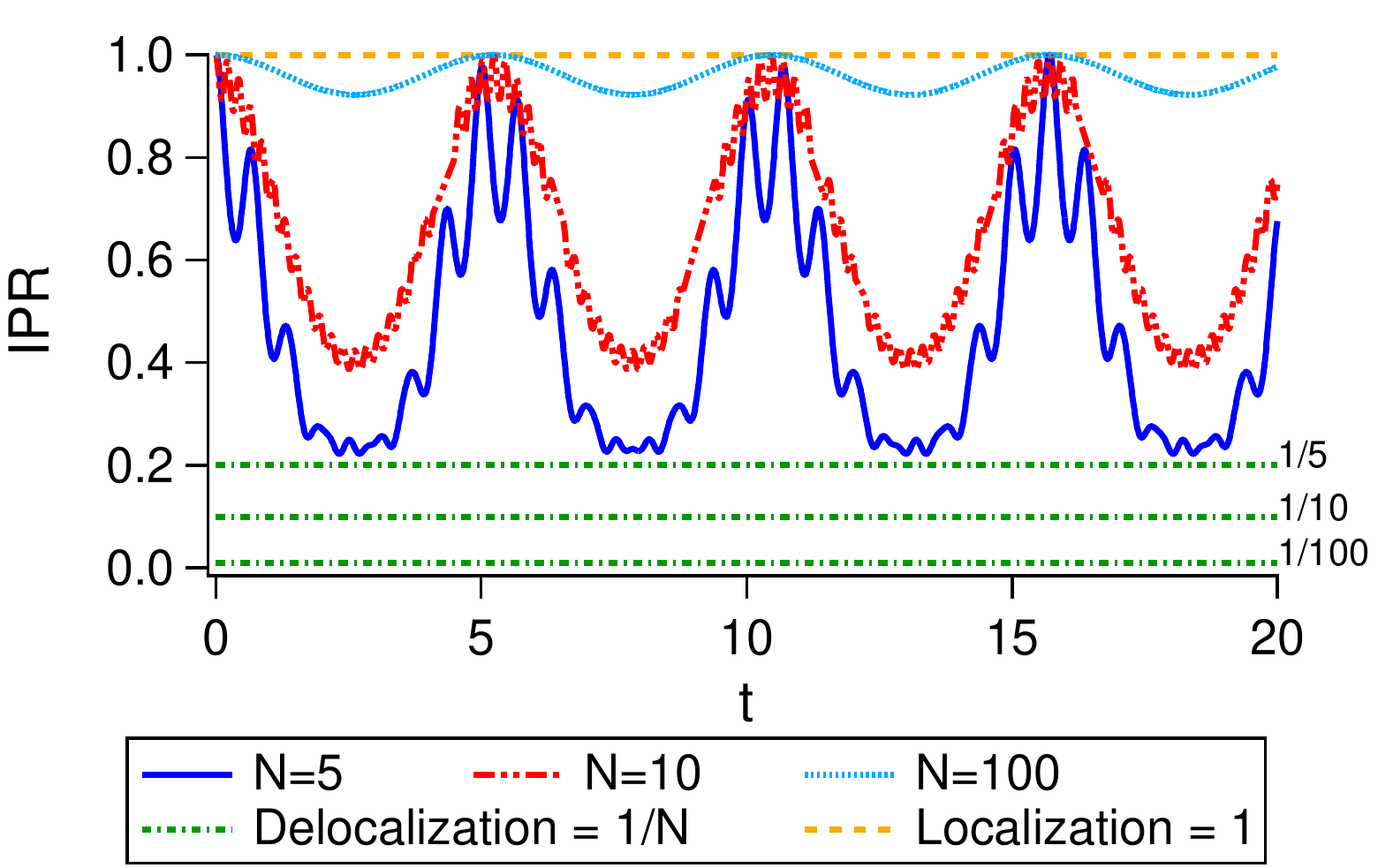}
	\caption{Inverse participation ratio (IPR) for a walker initially localized in $\ket{1}$ in the star graph. Results suggest that for $t>0$ there are instants of time when the IPR is close to the upper bound $1$ (orange dashed line), i.e. the localization. In particular, the IPR periodically reaches $1$ when the probability distribution is periodic. For $N>4$ the IPR does not reach the value $1/N$ (green dashdotted line), i.e. the delocalization. For $N\to+\infty$ the IPR approaches 1, since the probability of finding the walker in the starting vertex approaches $1$ (see Eqs. \eqref{eq:p_10_sg}--\eqref{eq:p_1i_sg}). Results for $\lambda=0.2$.}
	\label{fig:sg_IPR_multi}
\end{figure}

Next, we numerically evaluate the IPR \eqref{eq:ipr_local} for the probability distribution in Eqs. \eqref{eq:p_10_sg}--\eqref{eq:p_1i_sg}, and the results are shown in Fig. \ref{fig:sg_IPR_multi}. The IPR oscillates between $1$ and its minimum value, which grows with $N$, similarly to what happens in the complete graph. Indeed, for $N\to+\infty$ the IPR approaches 1 (localization), since the probability of finding the walker in the starting vertex approaches $1$. The periodicity of the IPR relies upon that of the probability distribution. When the latter is periodic, the IPR periodically reaches $1$, since the walker is initially localized in a vertex, and periodically comes back to it.  By considering $P_1(0,t\vert \lambda)=1/N$, we notice that the instantaneous exactly uniform mixing is never achievable for $N>4$, and so the IPR is never close to $1/N$, independently of $\lambda$. Instead, for $N \leq 4$ the mixing properties strongly depend on the choice of $N$ and $\lambda$, e.g. it is achievable for $\lambda=-1/(N+1)$ and for $N=2 \wedge \lambda=-1$. The instantaneous exactly uniform mixing is never achievable for $\lambda^\ast$, since $P_1(0,t\vert \lambda^\ast)=0 \,\forall\, t$.

Finally, we focus on the time dependence of the coherence of a walker initially localized in $\ket{1}$, which we derive in Appendix \ref{subapp:an_res_sg} and it is shown in Fig. \ref{fig:coher_star_loc}. The coherence shows a complex structure of local maxima and minima. However, it is smoother and periodic for the values of $\lambda$ which make the overall probability distribution periodic (see Appendix \ref{subapp:an_res_sg}), e.g. $\lambda=-1,-1/N,-1/(N+1)$.

\begin{figure}[htb]
	\centering	
	\includegraphics[width=0.45\textwidth]{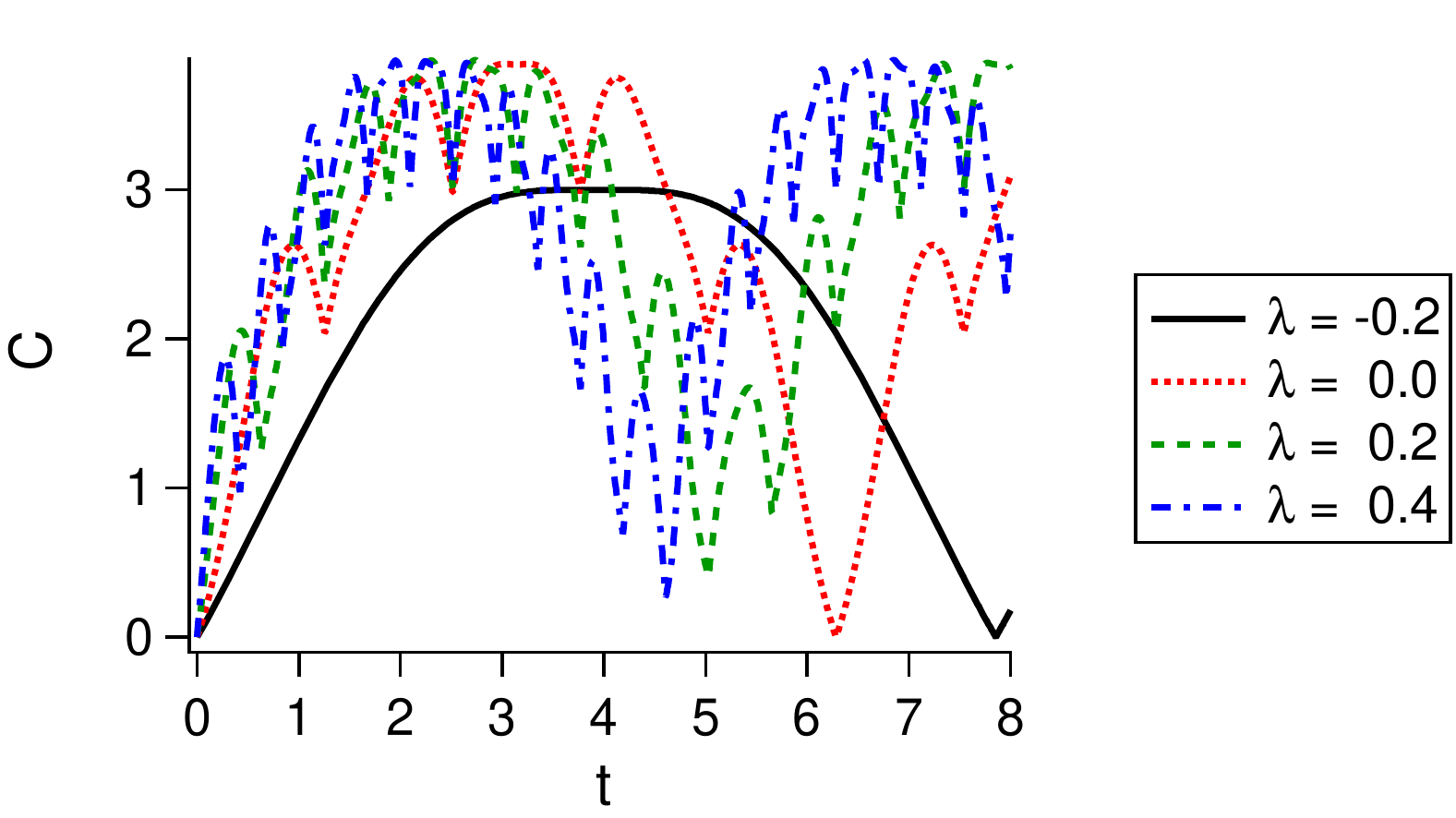}
	\caption{Coherence for a walker initially localized in $\ket{1}$ in the star graph with $N=5$. The coherence is smooth and periodic for $\lambda^\ast =-1/N$.}
	\label{fig:coher_star_loc}
\end{figure}

\section{Characterization}
\label{sec:estimation}
In this section we address the characterization of the CTQW Hamiltonian \eqref{eq:H_ctqwp4}, i.e. the estimation of the parameter $\lambda$ that quantifies the amplitude of the perturbation $\mathcal{H}_1=L^2$. Our aim is to assess whether, and to which extent, we may determine the value of $\lambda$ using only a snapshot of the walker dynamics, i.e. by performing measurements at a given time $t$. Hence, we briefly review some useful concepts from classical and quantum estimation theory \cite{paris2009quantum}.

Purpose of classical estimation theory is to find an {\em estimator}, i.e 
a function $\hat{\lambda}$ that, taking as input $n$ experimental data $\{x_i\}_{i=1,\dots,n}$ whose probabilistic distribution $P(x_i \vert \lambda)$ depends on $\lambda$, gives the most precise 
estimate of the parameter.  
A particular class of $\hat{\lambda}$ are the {\em unbiased estimators}, 
for which the expectation value is the actual value of the parameter 
$\lambda$, 
i.e. $E_\lambda [\hat{\lambda}] = \int dx P(x\vert \lambda) \hat{\lambda}(x) \equiv \lambda$. The main result regarding the precision of an estimator $\hat{\lambda}$ is given 
by the Cram\'{e}r-Rao Bound, which sets a lower bound on the variance of any 
unbiased estimator $\hat{\lambda}$, provided that the family of distribution 
$P(x\vert \lambda)$ realizes a so-called {\em regular statistical model}. In this case, the variance of any 
unbiased estimator $\hat{\lambda}$ satisfies the inequality
\begin{equation}
    \sigma^2(\hat{\lambda}) \geq \frac{1}{n \mathcal{F}_c(\lambda)}\,,
    \label{eq:classcrb}
\end{equation}
where $\mathcal{F}_c(\lambda)$ is the Fisher Information (FI) of the 
probability distribution $P(x\vert\lambda)$
\begin{equation}
    \mathcal{F}_c(\lambda) = \int dx \frac{(\partial_\lambda P(x\vert \lambda))^2}{P(x\vert \lambda)}\,.
    \label{eq:fi_def_cont}
\end{equation}
Regular models are those with a constant support, i.e. the region in which $P(x\vert \lambda) \neq 0$ does not depend on the parameter $\lambda$, and with non-singular FI. If these hypotheses are not satisfied, estimators with vanishing variance may be easily found. Optimal estimators are those saturating the inequality \eqref{eq:classcrb}, and it can be proved that for $n\to+\infty$ maximum likelihood estimators attain the lower bound \cite{newey1994large}.

In a quantum scenario, the parameter must be encoded in the density matrix 
of the system In turn, a {\em quantum statistical model} is defined as a family of quantum states $\{\rho_\lambda\}$ parametrized by the value of $\lambda$. In order to extract information from the system, we need to perform measurements, i.e. positive operator-valued measure (POVM) $\{\mathcal{E}_m\}$, where $m$ is a continuous or discrete index labeling the outcomes. According to the Born rule, a conditional distribution $P(m\vert \lambda) = \Tr{\rho_\lambda \mathcal{E}_m}$ naturally arises. Unlike the classical regime, the probability depends both on the state and on the measurement, so we can suitably choose them to get better estimates. In particular, given a family of quantum states $\left\{\rho_\lambda\right\}$, we can find a POVM which maximizes the FI, i.e.  
\begin{equation}
    \mathcal{F}_c (\lambda) \leq \mathcal{F}_q(\lambda) = \Tr{\rho_\lambda \Lambda_\lambda^2}\,,
    \label{eq:qfi_def}
\end{equation}
where $\mathcal{F}_q(\lambda)$ is the Quantum Fisher Information (QFI) and $\Lambda_\lambda$ is the Symmetric Logarithmic Derivative (SLD), which is implicitly defined as
\begin{equation}
    \frac{\rho_\lambda \Lambda_\lambda + \Lambda_\lambda \rho_\lambda}{2} = \partial_\lambda \rho_\lambda\,.
\end{equation}
The optimal POVM saturating the inequality \eqref{eq:qfi_def} is given by the projectors on the eigenspaces of the SLD. Since $\mathcal{F}_q(\lambda) = \max_{\mathcal{E}_m}\{\mathcal{F}_c(\lambda)\}$, we have a more precise bound on $\sigma^2(\hat{\lambda})$ which goes by the name of Quantum Cram\'{e}r-Rao (QCR) inequality
\begin{equation}
    \sigma^2(\hat{\lambda}) \geq \frac{1}{n\mathcal{F}_q(\lambda)}\,.
    \label{eq:qcrb}
\end{equation}
This establishes the ultimate lower bound of the precision in estimating a 
parameter $\lambda$ encoded in a quantum state. Notice that the QCR is valid for {\em regular quantum statistical model}, i.e. families of quantum states 
made of density matrices with constant rank (i.e. the rank does not depend 
on the parameter) and leading to a nonsingular QFI \cite{bycr,bcr,vrank}.
\par
In the present work we focus on pure states subjected to the unitary evolution in Eq. \eqref{eq:t_evol_op}, i.e. $\ket{\psi_\lambda(t)}={\mathcal{U}}_\lambda(t)\ket{\psi(0)}$. For such states the QFI reads as
\begin{equation}
	\mathcal{F}_q(t,\lambda) = 4 \left[\langle \partial_\lambda \psi_\lambda(t) \vert \partial_\lambda \psi_\lambda(t) \rangle- \abs{\langle \psi_\lambda(t) \vert \partial_\lambda \psi_\lambda(t) \rangle}^2 \right]\,.
	\label{eq:qfi_def_braket}
\end{equation}
When dealing with CTQWs on a graph, a reasonable and significant measurement is the position one. For such a measurement the FI reads as
\begin{equation}
    \mathcal{F}_c(t,\lambda) = \sum_{k=0}^{N-1} \frac{\left(\partial_\lambda P(k,t \vert \lambda)\right)^2}{P(k,t\vert\lambda)}\,,
    \label{eq:fi_def}
\end{equation}
where $P(k,t\vert\lambda)$ is the conditional probability of finding the walker in the $k$-th vertex at time $t$ when the value of the parameter is $\lambda$.

When the perturbation $\mathcal{H}_1$ commutes with the unperturbed Hamiltonian $\mathcal{H}_0$ (which is our case, see Eq. \eqref{eq:H_ctqwp4}), the unitary time evolution simplifies to
\begin{equation}
    \mathcal{U}_\lambda(t)  = e^{-it\mathcal{H}_0}e^{-i t\lambda \mathcal{H}_1}\,.
\end{equation}
Then, the QFI has a simple representation in terms of the perturbation and of time. Indeed, if our probe $\vert \psi \rangle$ at time $t=0$  does not depend on $\lambda$ and undergoes the evolution $\mathcal{U}_\lambda(t)$, at a later time $t>0$ we can write
\begin{align}
    \mathcal{F}_q(t) &= 4 t^2 \big[ \langle \psi \vert \mathcal{H}_1^2 \vert \psi \rangle -  \langle \psi \vert \mathcal{H}_1 \vert \psi \rangle ^2 \big]\nonumber\\
    &=4t^2 \langle (\Delta \mathcal{H}_1)^2\rangle\,.
    \label{eq:qfi_quadratic_time}
\end{align}
since $\vert \partial_\lambda \psi_\lambda (t) \rangle = -it\mathcal{H}_1 \vert \psi_\lambda (t)\rangle$ when $[\mathcal{H}_0,\mathcal{H}_1] = 0$. We emphasize that the QFI does not depend on the parameter $\lambda$ to be estimated. This is due to the unitary evolution, and to the fact that at $t=0$ the probe $\ket{\psi}$ does not depend on $\lambda$.

In the following, we evaluate the QFI of localized states, whose dynamics is addressed in Sec. \ref{sec:dynamics}, and we determine the states maximizing the QFI for cycle, complete, and star graph. We compare the QFI with the FI for a position measurement to assess whether it is an optimal measurement or not. Moreover, we find the simple graphs allowing the maximum QFI. Please refer to Appendix \ref{app:q_fi_scg} for the details about the analytical derivation of the results shown in the following.

\subsection{Localized states}

\subsubsection{Cycle graph}
\begin{figure}[htb]
	\centering	
	\includegraphics[width=0.45\textwidth]{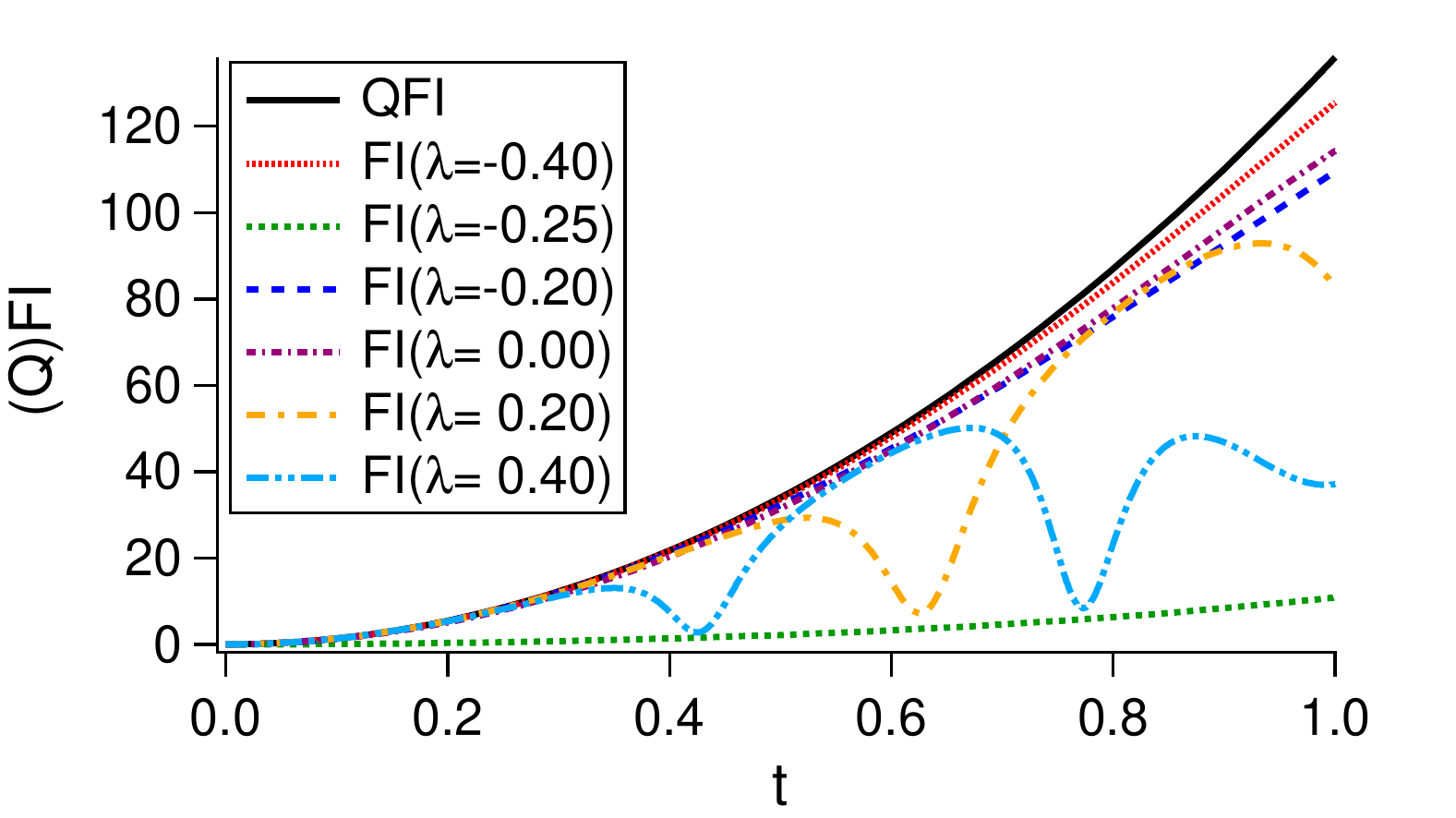}
	\caption{Quantum (black solid line) and classical Fisher Information (colored non-solid lines) of position measurement for an initially localized state in the cycle graph. Results for $N=5$.}
	\label{fig:qfi_cyc_loc}
\end{figure}

The QFI of an initially localized state in the cycle graph is
\begin{equation}
    \mathcal{F}_q (t)=136 t^2\,,
    \label{eq:qfi_loc_cyc}
\end{equation}
and it is independent of $N$. We numerically evaluate the FI \eqref{eq:fi_def} for the probability distribution in Eq. \eqref{eq:pjk_cyc_cos}. The results are shown in Fig. \ref{fig:qfi_cyc_loc} and suggest that the FI never reaches the QFI. Specific behaviors of the FI strongly depend on the choice of $N$ and $\lambda$.

\subsubsection{Complete Graph}
The QFI of an initially localized state in the complete graph is
\begin{equation}
    \mathcal{F}_q(N,t) = 4 N^2(N-1)t^2  \,.
    \label{eq:qfi_loc_cg}
\end{equation}
The FI is
\begin{equation}
    \mathcal{F}_c(N,t,\lambda) = \frac{4N^4(N-1) t^2\cos^2\left(\omega_N(\lambda) t \right)}{N^2-4(N-1) \sin^2\left(\omega_N(\lambda) t\right)}\,,
    \label{eq:fi_loc_cg}
\end{equation}
with $\omega_N(\lambda)$ defined in Eq. \eqref{eq:ang_freq}. Due to the symmetry of the graph, both the QFI and the FI do not depend on the starting vertex, i.e. the estimation is completely indifferent to the choice of the initially localized state. Unlike the QFI, the FI does depend on $\lambda$ and is symmetric with respect to $\lambda^\ast = -1/N$, as well as the probability distribution in Eqs. \eqref{eq:p0_cg}--\eqref{eq:pi_cg}. In particular, $\mathcal{F}_c(t,\lambda^\ast)=\mathcal{F}_q(t)$. {However, we recall that $P_0(0,t\vert\lambda^\ast) = 1$ and $P_0(i,t\vert\lambda^\ast) = 0$, i.e. the walker is in the starting vertex all the time. In this case the hypotheses leading to the 
Cram\'{e}r-Rao Bound \eqref{eq:classcrb} do not hold, since the model is
 not regular, and the bound may be easily surpassed. Indeed, if we perform 
 the measurement described by the POVM $\{\vert 0 \rangle \langle 0 \vert , \mathbbm{1}-\vert 0 \rangle \langle 0 \vert\}$, the variance of the estimator is identically zero, outperforming both classical and quantum bounds.}

\begin{figure}[htb]
	\centering	
	\includegraphics[width=0.45\textwidth]{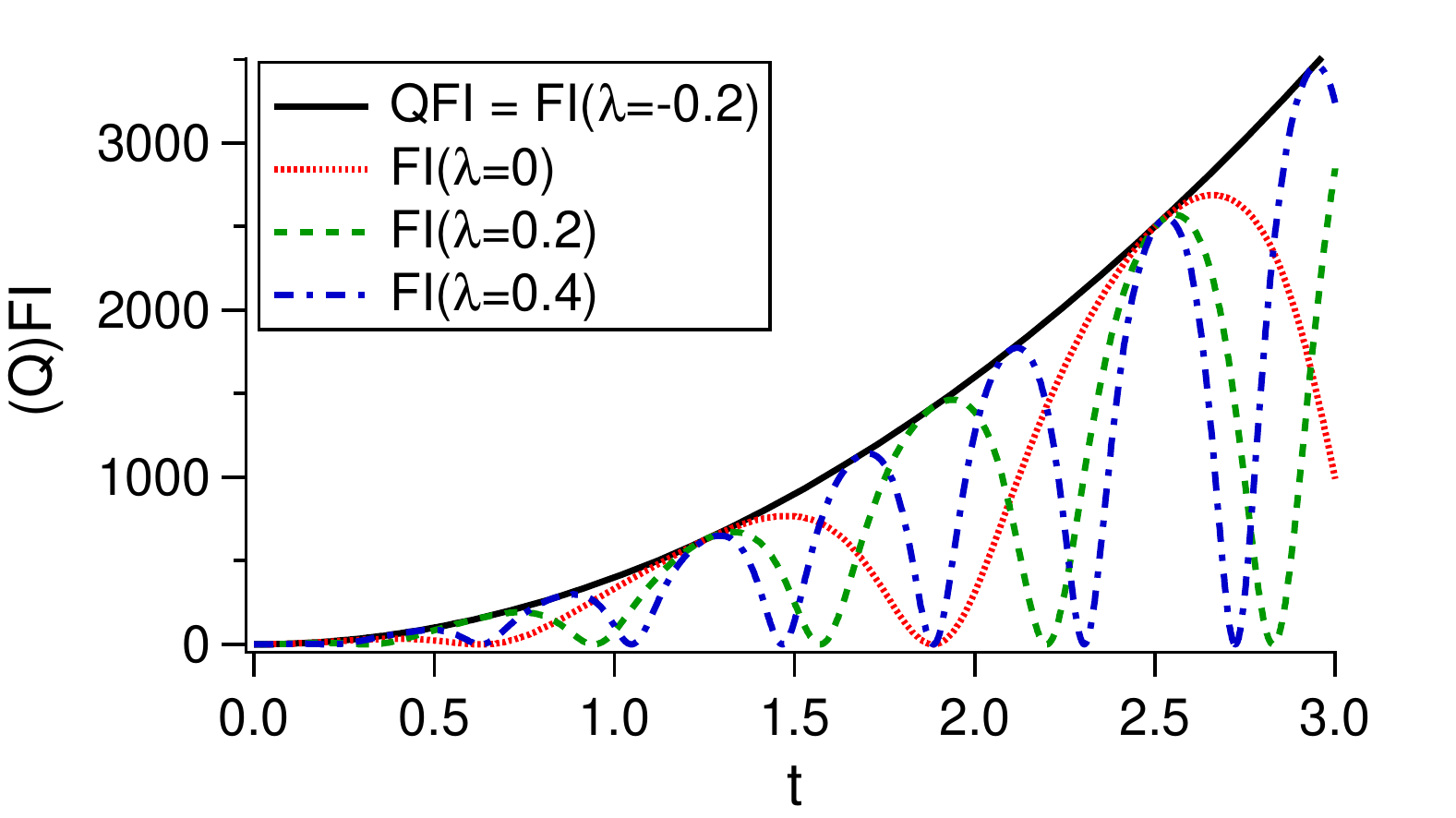}
	\caption{Quantum (black solid line) and classical Fisher Information (colored non-solid lines) of position measurement for an initially localized state in the complete graph. The same results are obtained for a walker initially localized in the central vertex $\ket{0}$ of the star graph of the same size. %The FI is optimal for $\lambda^\ast =-1/N$.
	Results for $N=5$.}
	\label{fig:qfi_cg_loc}
\end{figure}

For $\lambda \neq \lambda^\ast$, the periodicity of the probabilities in Eqs. \eqref{eq:p0_cg}--\eqref{eq:pi_cg} results in a dependence of the FI on $\lambda$ and an analogous oscillating behavior (Fig. \ref{fig:qfi_cg_loc}). The FI reaches periodically its local maxima when the numerator is maximum and the denominator is mininum, and these maxima saturate the Quantum Cram\'{e}r-Rao Bound
\begin{equation}
    \mathcal{F}_c(t_k,\lambda) = \mathcal{F}_q(t_k,\lambda)\,.
\end{equation} 
This occurs for $t_k= 2 k \pi /(N+\lambda N^2)$, with $k\in\mathbb{N}$, i.e. when the walker is completely localized and we definitely find it in the starting vertex. Indeed, in the probability distribution the parameter $\lambda$ is encoded only in the angular frequency, thus knowing when the walker is certainly in the starting vertex means knowing exactly its period, and thus the parameter $\lambda$.  
However, to perform such a measurement one needs some \textit{a priori} knowledge of the value of the parameter. In fact, the POVM saturating the Quantum Cram\'{e}r-Rao Bound \eqref{eq:qcrb} strongly depends on the parameter $\lambda$.

\subsubsection{Star Graph}
The time evolution of the state localized in the center of the star graph is equivalent to that of a localized state in the complete graph, as already pointed out in Sec. \ref{subsec:stargraph}. Thus, for this state the QFI and FI are provided in Eq. \eqref{eq:qfi_loc_cg} and Eq. \eqref{eq:fi_loc_cg}, respectively (see also Fig. \ref{fig:qfi_cg_loc}).

\begin{figure}[htb]
	\centering	
	\includegraphics[width=0.45\textwidth]{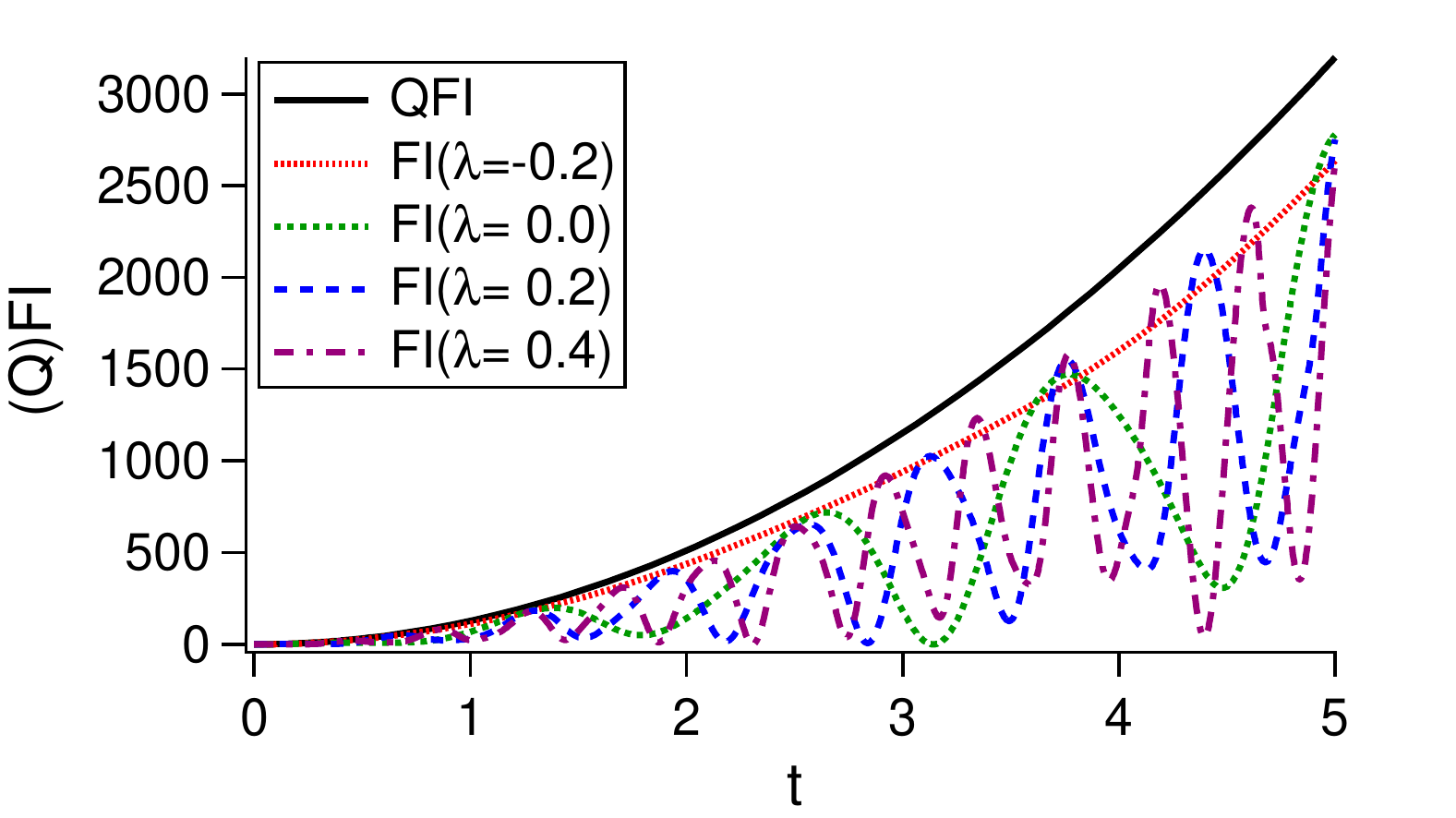}
	\caption{Quantum (black solid line) and classical Fisher Information (colored non-solid lines) of position measurement for a walker initially localized in an outer vertex of the star graph. Results for $N=5$.}
	\label{fig:qfi_sg_loc}
\end{figure}

Things change when we consider a walker initially localized in one of the outer vertices of the star graph. In this case the QFI is
\begin{equation}
	\mathcal{F}_q(N,t) =  4 (N^2+N-2) t^2\,.
	\label{eq:qfi_loc_sg}
\end{equation}
 We numerically evaluate the FI \eqref{eq:fi_def} for the probability distribution in Eqs. \eqref{eq:p_10_sg}--\eqref{eq:p_1i_sg} and the results are shown in Fig. \ref{fig:qfi_sg_loc}. Unlike the complete graph, for the star graph there is no saturation of the Quantum Cram\'{e}r-Rao Bound. Notice, however, that for $\lambda^\ast=-1/N$ the walker cannot reach the central site and, in principle, one may exploit this feature to build a non regular model, as we discussed in the previous section.
%%%%

\subsection{States maximizing the QFI}
In the previous section, we have studied how localized states behave as quantum probes for estimating the parameter $\lambda$ of the perturbation. However, we might be interested in finding the best estimate for such parameter by searching for the state $\rho_\lambda$ maximizing the QFI, hence minimizing the variance $\sigma^2(\hat{\lambda})$. For this purpose, it is worth introducing an alternative formula for QFI. When there is only one parameter to be estimated and the state is pure, the QFI reads as 
\begin{equation}
\mathcal{F}_q(\lambda,t)=\lim_{\delta\lambda\to0}\frac{8\left( 1-\abs{\braket{\psi_\lambda(t)}{\psi_{\lambda+\delta\lambda}(t)}}\right )}{\delta\lambda^2}\,.
\label{eq:qfi_def_dyn}
\end{equation}
This expression involves the modulus of the following scalar product:
\begin{equation}
\braket{\psi_\lambda(t)}{\psi_{\lambda+\delta\lambda}(t)}=\matrixel{\psi(0)}{{U}_{\delta\lambda}(t)}{\psi(0)}\,,
\label{eq:scpr}
\end{equation}
where 
\begin{align}
{U}_{\delta\lambda}(t)&:=e^{+i({\mathcal{H}}_0+\lambda{\mathcal{H}}_1)t}e^{-i[{\mathcal{H}}_0+(\lambda+\delta\lambda){\mathcal{H}}_1]t}\nonumber\\
&=e^{-i\delta\lambda{\mathcal{H}}_1t}
\label{eq:unitary_op}
\end{align}
is a unitary operator given by the product of two unitary operators \eqref{eq:t_evol_op} related to the time evolutions for $\lambda$ and $\lambda+\delta\lambda$, and the last equality holds since $[{\mathcal{H}}_0,{\mathcal{H}}_1]=0$ (see Eq. \eqref{eq:H_ctqwp4}).

The QFI strongly depends on the quantum state considered. To maximize the QFI, we recall the following lemma \cite{parthasarathy2001consistency}.
\begin{lemma}[K. R. Parthasarathy] Let $W$ be any unitary operator in the finite dimensional complex Hilbert space $\mathscr{H}$ with spectral resolution $\sum_{j=1}^k e^{i\theta_j}P_j$ where $e^{i\theta_1},\ldots,e^{i\theta_k}$ are the distinct eigenvalues of $W$ with respective eigenprojections $P_1,\ldots,P_k$. Define
\begin{equation}
m(W)=\min_{\norm{\psi}=1}\abs{\matrixel{\psi}{W}{\psi}}^2\,.
\end{equation}
Then the following hold:
\begin{enumerate}[(i)]
\item If there exists a unit vector $\ket{\psi_0}$ such that $\matrixel{\psi_0}{W}{\psi_0}=0$, then $m(W)=0$.
\item If $\matrixel{\psi}{W}{\psi}>0$ for every unit vector $\ket{\psi}$, then
\begin{equation}
m(W)=\min_{i\neq j}\cos^2\left ( \frac{\theta_i-\theta_j}{2}\right )\,.
\label{eq:mW_def}
\end{equation}
Furthermore, when the right-hand side is equal to $\cos^2\left ( \frac{\theta_{i_0}-\theta_{j_0}}{2}\right )$,
\begin{equation}
m(W)=\abs{\matrixel{\psi_0}{W}{\psi_0}}^2
\end{equation}
where
\begin{equation}
\ket{\psi_0} = \frac{1}{\sqrt{2}}\left (\ket{e_{i_0}}+\ket{e_{j_0}}\right )
\label{eq:psi_0}
\end{equation}
and $\ket{e_{i_0}}$ and $\ket{e_{j_0}}$ are arbitrary unit vectors in the range of $P_{i_0}$ and $P_{j_0}$
respectively.
\end{enumerate}
\label{lemma:parthasarathy}
\end{lemma}

The idea is to exploit the Lemma to compute the QFI. We consider $\ket{\psi_0}$ as initial state and we identify $W$ with ${U}_{\delta\lambda}(t)$, since $\braket{\psi_\lambda(t)}{\psi_{\lambda+\delta\lambda}(t)}=\langle \psi_0 \vert {U}_{\delta\lambda}(t)\vert \psi_0\rangle$, so that
\begin{equation}
\mathcal{F}_q(\lambda,t)=\lim_{\delta\lambda\to 0}\frac{8\left[ 1-\sqrt{m({U}_{\delta\lambda}(t))}\right]}{\delta\lambda^2}\,.
\label{eq:qfi_mw}
\end{equation}
Indeed, the state $\ket{\psi_0}$ in Eq. \eqref{eq:psi_0} maximizes the QFI by minimizing the modulus of the scalar product \eqref{eq:scpr}. The unit vectors involved by $\ket{\psi_0}$ are eigenvectors of the unitary operator \eqref{eq:unitary_op} and so, ultimately, of ${\mathcal{H}}_1$. In particular, such states are those whose eigenvalues minimize Eq. \eqref{eq:mW_def}.
The eigenvalues of the unitary operator \eqref{eq:unitary_op} are $e^{i\theta_j}=e^{-i\delta\lambda t \varepsilon_j^2}$, with $\{\varepsilon_j^2\}$ eigenvalues of ${\mathcal{H}}_1=\mathcal{H}_0^2$, being $\{\varepsilon_j\}$ those of ${\mathcal{H}}_0=L$. Thus, we can identify $\theta_j=-\delta\lambda t \varepsilon_j^2$. 
Because of this relation, we may assume $\ket{e_{i_0}}$ and $\ket{e_{j_0}}$ to be the eigenstates corresponding to the lowest and the highest energy eigenvalue. Indeed, in the limit for $\delta\lambda t \to 0$ the cosine in Eq. \eqref{eq:mW_def} is minimized by maximizing the difference $\theta_i-\theta_j$. Then, the QFI reads as follows:
\begin{equation}
\mathcal{F}_q(t) = t^2(\varepsilon_{max}^2-\varepsilon_{min}^2)^2=t^2 \varepsilon_{max}^4\,.
\label{eq:qfi_delta_eigenval}
\end{equation}
Because of the choice of the state $\ket{\psi_0}$, which involves the lowest and the highest energy eigenstates, the first equality follows from Eq. \eqref{eq:qfi_quadratic_time}, whereas the second equality holds since $\varepsilon_{min}=0$ for simple graphs. An eventual phase difference between the two eigenstates in Eq. \eqref{eq:psi_0} would result in the same QFI, but a different FI, as shown in Appendix \ref{subapp:phi_maxQFI_states}.

%%%% NEW ENTRY

\subsubsection{General Graph}
\label{subsubsec:general_graph_maxQFI}
We prove that for a specific class of graphs the maximum QFI is always equal to $N^4t^2$, provided that the probe of the system is the state \eqref{eq:psi_0}. Indeed, according to Lemma \ref{lemma:parthasarathy}, in order to find quantum probes maximizing the QFI, we need to search for systems whose eigenvalues separation is maximum. For a graph of $N$ vertices with no loops, the row sums and the column sums of the graph Laplacian $L_N$ are all equal to $0$, and the vector $(1,\ldots,1)$ is always an eigenvector of $L$ with eigenvalue $0$. It follows that any Laplacian spectrum contains the zero 
eigenvalue and to maximize the QFI we need to find graphs having the largest maximum eigenvalue.

Following Ref. \cite{brouwer2011spectra}, the Laplacian spectrum of a graph $G(V,E)$ is the set of the eigenvalues of $L_N$
\begin{equation}
    S_L(G) =\{\mu_1 = 0,\mu_2 ,\dots,\mu_N\} \,,
\end{equation}
where the eigenvalues $\mu_i$ are sorted in ascending order. To study the maximum eigenvalue $\mu_N$ we introduce the complementary graph $\bar{G}$ of $G$. The complementary graph $\bar{G}$ is defined on the same vertices of $G$ and two distinct vertices are adjacent in $\bar{G}$ if and only if they are not adjacent in $G$. So, the adjacency matrix $\bar{A}$ can be easily obtained from $A$ by replacing all the off-diagonals $0$s with $1$s and all the $1$s with $0$s. In other words
\begin{equation}
    \bar{A}_N= \mathbb{J}_N - \mathbbm{1}_N - A_N\,,
\end{equation}
where $\mathbb{J}_N$ denotes the $N\times N$ all-ones matrix and $\mathbbm{1}_N$ the $N\times N$ identity matrix. A vertex in $G$ can be at most adjacent to $N-1$ vertices, since no loops are allowed. Then, the degree $\bar{d}_j$ of a vertex in $\bar{G}$ is $N-1-d_j$, i.e. the complement to $N-1$ of the degree of the same vertex in $G$. The diagonal degree matrix is therefore
\begin{equation}
    \bar{D}_N = (N-1)\mathbbm{1}_N - D_N\,.
\end{equation}
In conclusion, the Laplacian matrix $\bar{L}_N$ associated to the complementary graph $\bar{G}$ is
\begin{equation}
\label{barlapl}
    \bar{L}_N = \bar{D}_N-\bar{A}_N =  N \mathbbm{1}_N - \mathbb{J}_N- L_N\,.
\end{equation}

\begin{lemma}
Any eigenvector $\vec{n}$ of $L_N$ is an eigenvector of $\bar{L}_N$. If the eigenvalue of $\vec{n}$ for $L_N$ is $0$, then it is $0$ also for $\bar{L}_N$. If the eigenvalue of $\vec{n}$ for $L_N$ is $\mu_i$, then the eigenvalue for $\bar{L}_N$ is $N - \mu_i$. Thus, the spectrum of $\bar{L}_N$ is given by 
\begin{equation}
    S_{\bar{L}}(\bar{G}) = \{0,N-\mu_N,\ldots,N-\mu_2\}\,,
    \label{eq:s_barL}
\end{equation}
where the eigenvalues are still sorted in ascending order.
\end{lemma}

Any $L_N$ is positive-semidefinite, i.e. $\mu_i \geq 0 \ \forall \ i$, so this holds for $\bar{L}_N$ too. According to these remarks and to  Eq. \eqref{eq:s_barL}, we then observe that $\mu_N \leq N$, i.e. the largest eigenvalue is bounded from above by the number of vertices $N$. 
Moreover, the second-smallest eigenvalue $\mu_2$ of $L_N$ is the algebraic connectivity of $G$: it is greater than 0 if and only if $G$ is a connected graph. Indeed, the algebraic multiplicity of the eigenvalue 0 is the number of connected components of the graph \cite{fiedler1973algebraic,mohar1991laplacian,marsden2013eigenvalues}. So, if $\bar{G}$ has at least two distinct components, then the second-smallest eigenvalue of $\bar{L}_N$ is $N-\mu_N=0$, from which $\mu_N=N$.

\begin{lemma}
    Given a graph $G$ and its Laplacian spectrum $S_{L}(G) = \{0,\mu_2,\ldots,\mu_N\}$, the largest Laplacian eigenvalue $\mu_N$ is bounded from above by $\mu_N \leq N$, and the equality is saturated only if the complementary graph $\bar{G}$ is disconnected.
\end{lemma}

This result in spectral graph theory has a direct impact on our estimation problem. Since our perturbation is the square of the graph Laplacian, the maximum QFI is given by Eq. \eqref{eq:qfi_delta_eigenval} and involves the lowest and the largest eigenvalue of the Laplacian spectrum. 
\begin{lemma}
    The simple graphs $G$ whose complementary graph $\bar{G}$ is disconnected are the only ones providing the maximum QFI for the estimate of the parameter $\lambda$ in Eq. \eqref{eq:H_ctqwp4}. For such graphs, the largest eigenvalue of the graph Laplacian is $N$ and the lowest is $0$. This results in the following maximum QFI 
    \begin{equation}
    \label{eq:maxQFIgraph}
        \mathcal{F}^{max}_q(N,t) = N^4 t^2 \,.
    \end{equation}
    \label{lemma:mleg_maxqfi}
\end{lemma}

This lemma allows us to predict whether or not a graph provides the maximum QFI and its value, with no need to diagonalize the graph Laplacian. Some graphs satisfying Lemma \ref{lemma:mleg_maxqfi} are the complete, the star, the wheel, and the complete bipartite graph. The cycle graph allows the maximum QFI only for $N\leq 4$: for $N=2,3$ it is just a complete graph; for $N=4$ the complementary graph has two disconnected components, and for $N>4$ it is connected.

\subsubsection{Cycle graph}
\label{subsubsec:cycle_max_qfi}
\begin{figure}[htb]
	\centering	
	\includegraphics[width=0.45\textwidth]{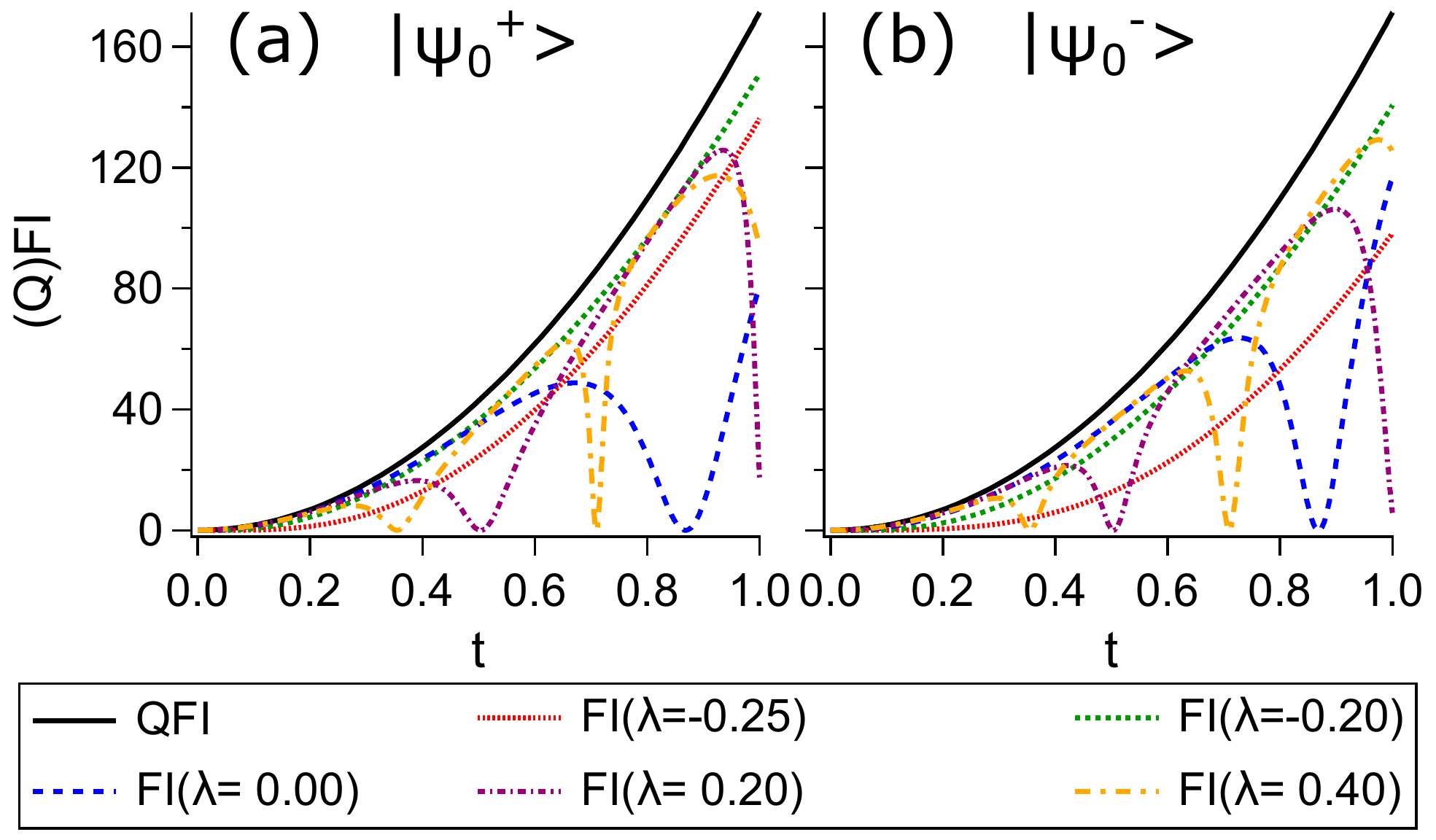}
	\caption{Quantum (black solid line) and classical Fisher Information (colored non-solid lines) of position measurement for the states maximizing the QFI in the cycle graph for odd $N$: (a) $\ket{\psi_0^+}$, where the highest energy state is Eq. \eqref{eq:evec_cyc_max_cos}; and (b) $\ket{\psi_0^-}$, where the highest energy state is Eq. \eqref{eq:evec_cyc_max_sin}. For odd $N$, indeed, the highest energy level is doubly degenerate. While the QFI does not depend on the choice of the corresponding eigenstate, the FI does. %In particular, the FI is equal to the QFI when we choose the highest energy state by definition (Table \ref{tab:eig_pbm_cycle}). Instead, when we choose Eq. \eqref{eq:evec_cyc_max_cos} or \eqref{eq:evec_cyc_max_sin}, the FI never reaches the QFI.
	Results for $N=5$.}
	\label{fig:qfi_cyc_max}
\end{figure}

The cycle graph satisfies Lemma \ref{lemma:mleg_maxqfi} only for $N\leq 4$. For $N>4$ the maximum QFI is lower than $N^4t^2$, and it depends on $N$. Indeed, the energy spectrum of the cycle graph is sensitive to the parity of $N$, and the state maximizing the QFI at $t=0$ is therefore
\begin{equation}
\vert\psi_0^{(\pm)}\rangle= \frac{1}{\sqrt{2}}(\ket{e_{min}}+\vert e_{max}^{(\pm)}\rangle)\,,
\label{eq:psi0_Nevenodd}
\end{equation}
where $\vert e_{min} \rangle$ is the ground state, while $\vert e_{max}^{(\pm)} \rangle$ is the eigenstate corresponding to the highest energy level and it depends on the parity of $N$. For even $N$ it is unique, whereas for odd $N$ the highest energy level is doubly degenerate, which is the reason for the $\pm$ sign (see Eqs. \eqref{eq:evec_cyc_max_cos}--\eqref{eq:evec_cyc_max_sin} and Table \ref{tab:eig_pbm_cycle}).
% In particular, for even $N$ the optimal state reads as
% \begin{equation}
% \ket{\psi_0}=\sqrt{\frac{2}{N}}\sum_{k=0}^{N/2-1}\ket{2k}\,,
% \label{eq:psi0_evenN_even}
% \end{equation}
% i.e. a superposition of the position states corresponding to even vertices.
The resulting QFI is
\begin{align}
\mathcal{F}_q(t)=
\begin{cases}
256 t^2 & \text{if $N$ is even}\,,\\
16 \left[1+\cos\left (\frac{\pi}{N}\right )\right]^4 t^2 & \text{if $N$ is odd}\,.
\end{cases}
\label{eq:qfi_cycle_evenodd}
\end{align}
The QFI for odd $N$ depends on $N$, and for large $N$ it approaches the QFI for even $N$, which, instead, does not depend on $N$. Even the FI discriminates between even and odd $N$, because of the ambiguity in choosing the highest energy eigenstate for odd $N$ (see Appendix \ref{subapp:q_fi_cyc}). For even $N$ the position measurement is optimal, i.e. $\mathcal{F}_c(t)=\mathcal{F}_q(t)$.
For odd $N$, both the eigenstates for $n=(N\pm1)/2$ in Table \ref{tab:eig_pbm_cycle} lead to $\mathcal{F}_c(t)=\mathcal{F}_q(t)$. Instead, if we choose the linear combinations of them in Eqs. \eqref{eq:evec_cyc_max_cos}--\eqref{eq:evec_cyc_max_sin}, the FI of position measurement is no longer optimal, as shown in Fig. \ref{fig:qfi_cyc_max}.

\subsubsection{Complete graph}
\label{subsubsec:complete_max_qfi}
The complementary graph of the complete graph has $N$ disconnected components, so it satisfies the Lemma \ref{lemma:mleg_maxqfi}.
A possible choice of the state maximizing the QFI (at $t=0$) is the following
\begin{equation}
    \vert \psi_0^l \rangle = \frac{1}{\sqrt{2}} ( \vert e_0 \rangle + \vert e^l_1 \rangle)\,,
    \label{eq:max_qfi_states_cg}
\end{equation}
where $\vert e_0 \rangle$ is the ground state, while $\vert e_1^l \rangle$, with $l=1,\ldots,N-1$, is the eigenstate corresponding to the highest energy level $\varepsilon_1 = N$, which is $(N-1)$-degenerate (see Table \ref{tab:eig_pbm_complete}). Then, we are free to choose any eigenstate from the eigenspace $\{\vert e^l_1 \rangle\}$ (or even a superposition of them) and the QFI is always given by Eq. \eqref{eq:maxQFIgraph}. On the other hand, the FI does depend on the choice of $\vert e^l_1 \rangle$.

\begin{figure}[htb]
	\centering	
	\includegraphics[width=0.45\textwidth]{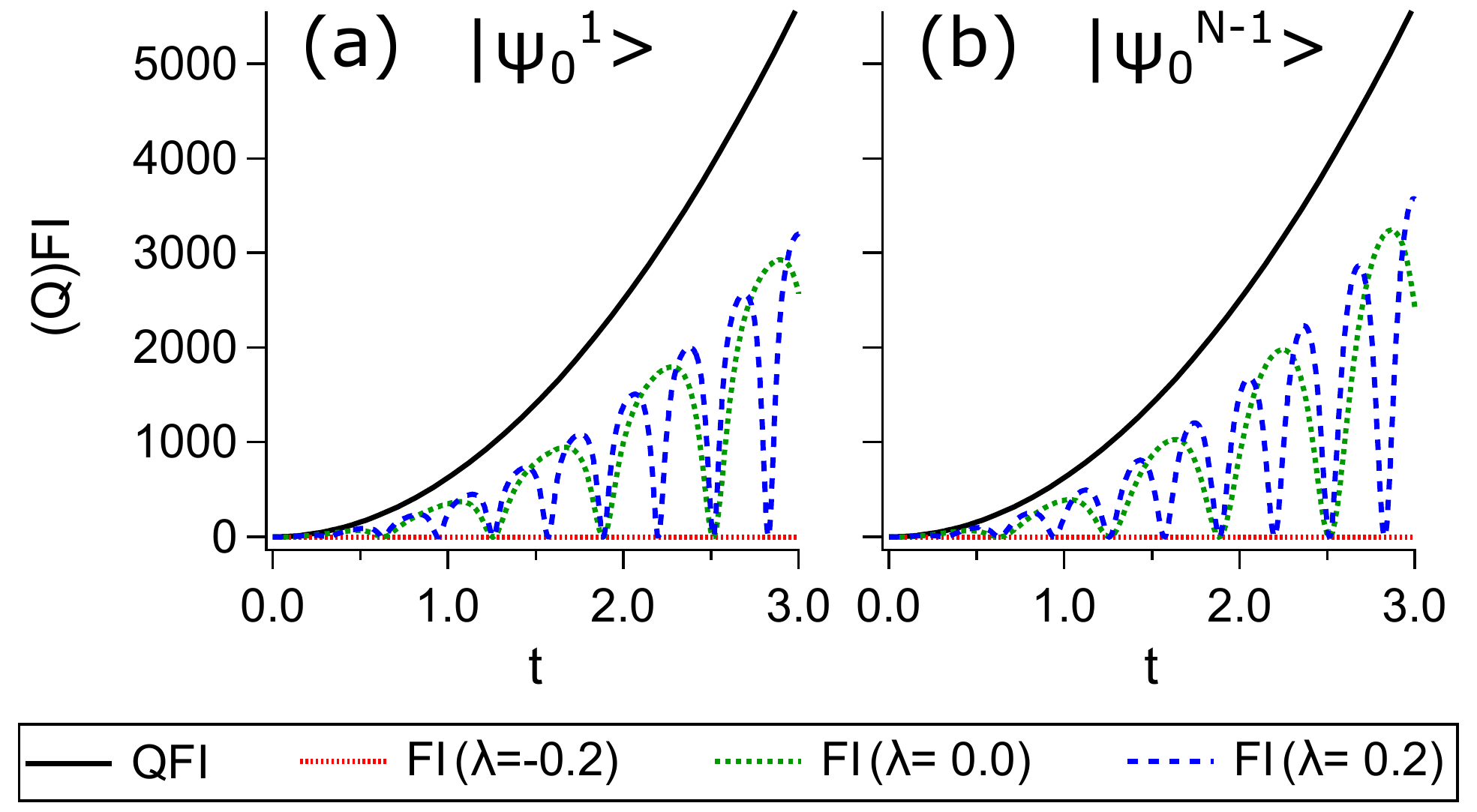}
	\caption{Quantum (black solid line) and classical Fisher Information (colored non-solid lines) of position measurement for two of the states maximizing the QFI in the complete graph: (a) $\ket{\psi_0^1}$ in Eq. \eqref{eq:max_qfi_states_cg_1} and (b) $\ket{\psi_0^{N-1}}$ in Eq. \eqref{eq:max_qfi_states_cg_2}. Due to the degeneracy of the highest energy level, there are several states providing the same maximum QFI. While the QFI does not depend on the choice of such states, the FI does. In both cases, the FI vanishes for $\lambda^\ast =-1/N$.  (b) also represents the results for the maximum QFI state in Eq. \eqref{eq:max_qfi_states_sg} for the star graph of the same size. Results for $N=5$.}
	\label{fig:qfi_cg_max}
\end{figure}

As an example, let us consider the two states
\begin{align}
    \vert \psi_0^1 \rangle &= \frac{1}{\sqrt{2}}(\vert e_0 \rangle + \vert e^1_1 \rangle )\,, \label{eq:max_qfi_states_cg_1}\\
    \vert \psi_0^{N-1} \rangle &= \frac{1}{\sqrt{2}}(\vert e_0 \rangle + \vert e^{N-1}_1 \rangle\,. \label{eq:max_qfi_states_cg_2}
\end{align}
These states are equivalent for the QFI (both maximize it), but they are not for the FI (see Fig. \ref{fig:qfi_cg_max}), which reads as follows 
\begin{align}
    &\mathcal{F}_c(\vert \psi_0^1\rangle;N,t,\lambda) = \frac{4 N^4 (N+2)t^2\sin^2(2t \omega_N(\lambda))}{(N+2)^2-8N \cos^2(2t\omega_N(\lambda))}\,, \label{eq:fi_max_qfi_states_cg_1}\\
    &\mathcal{F}_c(\vert \psi_0^{N-1} \rangle; N,t,\lambda) = \frac{4 N^4 (N-1) t^2 \sin ^2(2t\omega_N(\lambda))}{N^2-4 (N-1) \cos ^2(2t\omega_N(\lambda))}\,. \label{eq:fi_max_qfi_states_cg_2}
\end{align}
In both cases the FI is symmetric with respect to $\lambda^\ast=-1/N$, and for such value it vanishes. The local maxima occur for $t_k = \pi(k+1/2)/(N+\lambda N^2)$, with $k\in\mathbb{N}$, and are the following
\begin{align}
    \mathcal{F}^{max}_c(\vert \psi_0^1\rangle;N,t_k,\lambda) &= \frac{4 N^4}{N+2} t_k^2\,, \\
    \mathcal{F}^{max}_c(\vert \psi_0^{N-1} \rangle; N,t_k,\lambda)&= 4(N-1)N^2t_k^2\,. \label{eq:qfi_local_max_qfi_states_cg}
\end{align}
For these states the FI never reaches the value of the QFI \eqref{eq:maxQFIgraph}, so the position measurement on $\vert \psi^l_0 \rangle$ is not optimal.

\subsubsection{Star graph}
\label{subsubsec:star_max_qfi}
The complementary graph of the star graph has two disconnected components, so it satisfies the Lemma \ref{lemma:mleg_maxqfi}. The state maximizing the QFI (at $t=0$) is 
\begin{equation}
	\vert \psi_0 \rangle = \frac{1}{\sqrt{2}} \left(\vert e_0 \rangle + \vert e_2 \rangle\right)\,,
	\label{eq:max_qfi_states_sg}
\end{equation}
where $\vert e_0 \rangle$ is the ground state, while $\vert e_2 \rangle$ is the eigenstate corresponding to the highest energy level $\varepsilon_2 = N$ (see Table \ref{tab:eig_pbm_star}). The resulting QFI is given by Eq. \eqref{eq:maxQFIgraph}. Since the highest energy level is not degenerate, there is no ambiguity in the state maximizing the QFI. For such state the FI reads as Eq. \eqref{eq:fi_max_qfi_states_cg_2}, so please also refer to Fig. \ref{fig:qfi_cg_max}(b). 

\begin{table*}[tb]
    \centering
    \begin{ruledtabular}
        \begin{tabular}{rcccccc}
        & \multicolumn{3}{c}{\textbf{QFI}}  & \multicolumn{3}{c}{\textbf{FI}}\\ \cline{2-4}\cline{5-7}
        & \textit{cycle} & \textit{complete} & \textit{star} & \textit{cycle} & \textit{complete} & \textit{star}\\ \hline
        \textit{Localized states} & $\mathcal{O}(1)$ & $\mathcal{O}(N^3)$ & $\mathcal{O}(N^2)$ & $\mathcal{O}(1)$ & $\mathcal{O}(N^3)$ & $\mathcal{O}(N^2)$\\
        \textit{Maximum QFI states} & $\mathcal{O}(1)$ & $\mathcal{O}(N^4)$ & $\mathcal{O}(N^4)$ & $\mathcal{O}(1)$ & $\mathcal{O}(N^3)$ & $\mathcal{O}(N^3)$\\
        \end{tabular}
    \end{ruledtabular}
    \caption{Asymptotic behavior of the quantum Fisher information and of the classical Fisher information for large order $N$ of the cycle, complete, and star graphs, for localized and maximum QFI states.}
    \label{tab:big-o_q_fi}
\end{table*}

\begin{table*}[tb]
    \centering
    \begin{ruledtabular}
        \begin{tabular}{rccc}
        & \multicolumn{3}{c}{\textbf{FI}}\\ \cline{2-4}
        & \textit{cycle} & \textit{complete} & \textit{star}\\ \hline
        \textit{Localized states} &  $\mathcal{O}(t^2)$ & $\mathcal{O}(t^2)$ & $\mathcal{O}(t^2)$\\
        \textit{Maximum QFI states} & 
           \multicolumn{1}{l}{$\mathcal{O}(t^2)$ for energy eigenstates in Table \ref{tab:eig_pbm_cycle}}& $\mathcal{O}(t^4)$ & $\mathcal{O}(t^4)$\\
         &\multicolumn{1}{l}{$\mathcal{O}(t^4)$ for odd $N$ and highest energy eigenstate \eqref{eq:evec_cyc_max_cos} or \eqref{eq:evec_cyc_max_sin}}
        \end{tabular}
    \end{ruledtabular}
    \caption{Behavior at short times $t$ of the classical Fisher information of the cycle, complete, and star graphs, for localized and maximum QFI states. The maximum QFI state is the superposition of the ground state and the highest energy eigenstate. The QFI is always $\mathcal{O}(t^2)$, even at short and long times (see Eq. \eqref{eq:qfi_quadratic_time}), since the perturbation $\mathcal{H}_1$ is time-independent.}
    \label{tab:smallt_q_fi}
\end{table*}

\section{Discussion and Conclusions}
\label{sec:conclusions}
In this paper we have investigated the dynamics and the characterization of continuous-time quantum walks (CTQW) with Hamiltonians of the form $\mathcal{H}= L + \lambda L^2$, being $L$ the Laplacian (Kirchhoff) matrix of the underlying graph. We have considered cycle, complete, and star graphs, as they describe paradigmatic models with low/high connectivity and/or symmetry. The perturbation $\lambda L^2$ to the CTQW Hamiltonian $L$ introduces next-nearest-neighbor hopping. This strongly affects the CTQW in the cycle and in the star graph, whereas it is negligible in the complete graph, since each of its vertices is adjacent to all the others and $L^2=NL$. Clearly $[L,\lambda L^2]=0$, so the commutator between the unperturbed Hamiltonian and the perturbation is not indicative of how much the system is perturbed. Therefore, we consider how different is $L^2$ from $L$ by assessing the Frobenius norm of the operator $\Delta=L - L^2/N$, i.e. $\norm{\Delta}_F=\sqrt{\Tr{\Delta^\dagger \Delta}}$ \cite{horn2012matrix}. This turns out to be null for the complete graph, equal to $\sqrt{6N-40+70/N}$ for the cycle, and to $\sqrt{N-4+5/N-2/N^2}$ for the star graph. According to this, the cycle graph is the most perturbed, and the complete the least one.

Our results indicate the general quantum features of CTQWs on graphs, e.g. revivals, interference, and creation of coherence, are still present in their perturbed versions. On the other hand, novel interesting effects emerge, such as the appearance of symmetries in the behavior of the probability distribution and of the coherence. In the cycle graph (for $t\ll 1$), the perturbation affects the speed of the walker, while preserving the ballistic spreading. The variance is symmetric with respect to $\lambda _0$, despite the fact that the probability distribution is not. The value $\lambda_0$ makes the next-nearest-neighbor hopping equal to the nearest-neighbor hopping. The physical interpretation of this behavior is still an open question, which deserves to be further investigated. In the complete graph the perturbation does not affect the dynamics, since $L^2=N L$, so the resulting perturbed Hamiltonian is proportional to $L$. In the star graph, the perturbation affects the periodicity of the system. We have determined the values of $\lambda$ allowing the system to be periodic, thus to have exact revivals. In particular, the value $\lambda^\ast=-1/N$ makes the walker live only in the outer vertices, provided it starts in one of them.

Characterizing the perturbed Hamiltonian amounts to estimating the parameter $\lambda$ of the perturbation. We have addressed the optimal estimation of $\lambda$ by means of the quantum Fisher information (QFI) and using only a snapshot of the walker dynamics. The states maximizing the QFI turn out to be the equally-weighted linear combination of the eigenstates corresponding to the lowest and highest energy level. In addition, we have found that the simple graphs whose complementary graph is disconnected, e.g. the complete and the star graph, are the only ones providing the maximum QFI $N^4 t^2$. Moreover, we have evaluated the Fisher information (FI) of position measurements to assess whether it is optimal. We sum up the asymptotic behavior of the (Q)FI for large $N$ in Table \ref{tab:big-o_q_fi} and for $t\ll1$ in Table \ref{tab:smallt_q_fi}. When the probe is a localized state, the QFI in the cycle graph is independent of the order $N$ of the graph. In the complete graph, the local maxima of the FI equal the QFI and occur when the walker is localized in the starting vertex with probability $1$, and this happens periodically. However, to perform such a measurement one needs some \textit{a priori} knowledge of the value of $\lambda$. When the probe is the maximum QFI state, the QFI in the cycle graph depends on $N$, and FI is optimal for even $N$. In general, when the highest energy level is degenerate, the QFI does not depend on the choice of the corresponding eigenstates when defining the optimal state, instead the FI does.

Besides fundamental interest, our study may find applications in designing enhanced algorithms on graphs, e.g. spatial search, and as a necessary ingredient to study dephasing and decoherence.

\acknowledgments
This study has been partially supported by SERB through the 
VAJRA scheme (grant VJR/2017/000011). P.B. and M.G.A.P. are 
members of GNFM-INdAM.

A.C. and L.R. contributed equally to this work.

\appendix

\section{Analytical derivation of the results for the dynamics}
\label{app:an_res_ccsg}
The dynamics of the system is essentially encoded in the time evolution of the density matrix. For an initially localized state $\ket{i}$, the density matrix is given by $\rho(t) = \vert i(t) \rangle \langle i(t) \vert$, whose generic element in the position basis is
\begin{align}
    \rho_{j,k}(t) & =  \langle j \vert \mathcal{U}_\lambda(t) \vert i \rangle  \langle i\vert \mathcal{U}^\dag_\lambda(t) \vert k \rangle\,,
    \label{eq:rho_mn}
\end{align}
where the time-evolution operator $\mathcal{U}_\lambda(t)$ is defined in Eq. \eqref{eq:t_evol_op}. The probability distribution is given by the diagonal elements of the density matrix
\begin{equation}
    P_i(j,t\vert \lambda) = \vert \langle j \vert i(t) \rangle \vert^2 = \langle j \vert  i(t) \rangle \langle i(t) \vert j \rangle = \rho_{j,j}(t)\,.
    \label{eq:prob_from_state}
\end{equation}
On the other hand, the modulus of the off-diagonal elements of the density matrix entering the definition of coherence in Eq. \eqref{eq:coherence_def} can also be expressed in terms of probabilities:
\begin{align}
    \abs{\rho_{j,k}(t)} &=\vert \langle j \vert \mathcal{U}_\lambda(t) \vert i \rangle \vert \vert \langle k \vert \mathcal{U}_\lambda(t) \vert i \rangle \vert  \nonumber \\ &=\sqrt{P_i(j,t\vert\lambda)P_i(k,t\vert\lambda)}\,.
    \label{eq:abs_rho_mn}
\end{align}

\subsection{Cycle Graph}
\label{subapp:an_res_cyc}
According to the time-evolution operator and to the spectral decomposition in Table \ref{tab:eig_pbm_cycle}, in a cycle graph an initially localized state $\ket{j}$ evolves in time as follows
\begin{equation}
    \ket{j(t)} =\frac{1}{\sqrt{N}}\sum_{n=0}^{N-1}e^{-iE_n^\lambda t}e^{i\frac{2\pi }{N}jn}\ket{e_n}\,,
    \label{eq:jt_cyc}
\end{equation}
where $E_n^\lambda:=\varepsilon_n+\lambda \varepsilon_n^2$, and $\exp\{ i\frac{2\pi }{N}jn \} /\sqrt{N}=\braket{e_n}{j}$. Then, the probability of finding the walker in the vertex $k$ at time $t$ is
\begin{equation}
   P_j(k,t\vert\lambda) =\frac{1}{N^2}\sum_{n,m=0}^{N-1}e^{-i(E_n^\lambda-E_m^\lambda)t}e^{i\frac{2\pi }{N}(n-m)(j-k)}\,. \label{eq:pjk_cyc}
\end{equation}
This expression leads to Eq. \eqref{eq:pjk_cyc_cos} as follows. Let $p_{nm}$ be the summand, excluding $1/N^2$. The summation over $m$ can be split in three different summations: one over $m=n$ (providing $\sum_{n}p_{nn}=N$), one over $m>n$, and one over $m<n$. Since $p_{nm}=p_{mn}^\ast$, then  $\sum_{m<n}p_{nm}=\sum_{m>n}p_{mn}^\ast$, so $\sum_{m>n}(p_{nm}+p_{mn}^\ast)=2\sum_{m>n}\Re\{p_{nm}\}$, with $\Re\{p_{nm}\}=\cos[\operatorname{arg} (p_{nm})]$.

To prove that the probability distribution is symmetric with respect to the starting vertex $j$, i.e. that $P_j(j+k,t\vert \lambda)=P_j(j-k,t\vert \lambda)$, we consider Eq. \eqref{eq:pjk_cyc}. The left-hand side is
\begin{equation}
	P_j(j+k,t\vert \lambda) = \frac{1}{N^2}\sum_{n,m=0}^{N-1}e^{-i(E_n^\lambda-E_m^\lambda)t}e^{i\frac{2\pi }{N}(n-m)(-k)}\,.
\end{equation}
Now, letting $l = N-n$ ($q=N-m$) be the new summation indices, and since $\varepsilon_{N-l} = \varepsilon_l$ ($\varepsilon_{N-q} = \varepsilon_q$), we have that
\begin{align}
	P_j(j+k,t\vert \lambda) &= \frac{1}{N^2} \sum_{l,q=1}^{N}e^{-i(E_{l}^\lambda-E_{q}^\lambda)t}e^{i\frac{2\pi }{N}(q-l)(-k)} \nonumber \\
	&= \frac{1}{N^2} \sum_{l,q=0}^{N-1}e^{-i(E_{l}^\lambda-E_{q}^\lambda)t}e^{i\frac{2\pi }{N}(l-q)[j-(j-k)]} \nonumber \\
    &= P_j(j-k, t \vert \lambda)\,.
\end{align}
The second equality holds since the summand of index $l=N$ ($q=N$) is equal to that of index  $l=0$ ($q=0$). Indeed, according to Table \ref{tab:eig_pbm_cycle}, the virtual $E_N^\lambda$ is equal to $E_0^\lambda$ (the actual energies have index running from $0$ to $N-1$). In addition, $\exp\{i\frac{2\pi }{N}(q-l)k\}$ returns the same value if evaluated in $l=N$ ($q=N$) or $l=0$ ($q=0$).

Finally, we justify the expression of the variance of the position in Eq. \eqref{eq:xVar_cyc}. We assume even $N$, $\ket{j=N/2}$ as initial state, and $t\ll 1$. The variance requires the expectation values of $\hat{x}$ and $\hat{x}^2$, and the vertex states are eigenstates of the position operator. The probability distribution \eqref{eq:pjk_cyc_cos} is symmetric about the starting vertex, thus $\langle \hat{x} \rangle=N/2$, and it involves summands of the form
\begin{align}
\cos(\alpha t +\beta)&=\cos(\alpha t) \cos \beta -\sin( \alpha t) \sin\beta\nonumber\\
&= (1-\frac{\alpha^2}{2}t^2)\cos \beta-\alpha t \sin \beta+\mathcal{O}(t^3)\,,
\end{align}
since $t\ll 1$. Hence, letting $\alpha_{nm}=E_n^\lambda-E_m^\lambda$ and $\beta_{nm}^{kj}=\frac{2\pi}{N}(n-m)(k-j)$, we can write
\begin{align}
    \langle \hat{x}^2(t) \rangle&\approx\frac{1}{N}\sum_{k=0}^{N-1} k^2+\frac{2}{N^2}\sum_{\substack{n=0, \\ m>n}}^{N-1}\sum_{k=0}^{N-1}\left[\vphantom{\frac{t^2}{2}} k^2 \cos{\beta_{nm}^{kj}}\right.\nonumber\\
    &\quad\left.-\frac{t^2}{2} k^2 \alpha_{nm}^2 \cos{\beta_{nm}^{kj}}-t k^2 \alpha_{nm}  \sin{\beta_{nm}^{kj}}\right]\nonumber\\
    &=\frac{1}{6}(N-1)(2N-1)-\frac{1}{12}[N(N-6)+2]\nonumber\\
    &\quad+2t^2 (20\lambda^2+8\lambda+1)\nonumber\\
    &=\frac{N^2}{4}+\Big[ 40\Big(\lambda+\frac{1}{5}\Big)^2+\frac{2}{5}\Big]t^2\,.
\end{align}
Then, the variance \eqref{eq:xVar_cyc} follows, and is symmetric with respect to $\lambda_0 = -1/5$.

\subsection{Complete Graph}
\label{subapp:an_res_cg}
The complete graph has two energy levels (see Table \ref{tab:eig_pbm_complete}), thus the unitary time-evolution operator has the following spectral decomposition:
\begin{equation}
    \mathcal{U}_\lambda(t) =  \vert e_0\rangle \langle e_0 \vert + e^{-2i\omega_N(\lambda)t} \sum_{l=1}^{N-1} \vert e^l_1\rangle \langle e^l_1 \vert \,,
\end{equation}
being $\omega_N(\lambda)$ defined in Eq. \eqref{eq:ang_freq}. Hence, a localized state $\ket{0}$ evolves in time according to
\begin{align}
    \vert 0(t) \rangle = & \frac{1}{N}\left[1+(N-1)e^{-i2\omega_N(\lambda)t}\right] \vert 0 \rangle  \nonumber\\
    &+\frac{1}{N} \left(1-e^{-i2\omega_N(\lambda)t}\right)\sum_{k=1}^{N-1} \vert k \rangle\,.
    \label{eq:jt_cg}
\end{align}

Then, the density matrix $\rho(t) = \vert 0(t) \rangle \langle 0(t) \vert$ in the position basis is
\begin{align}
     \rho(t)= &[1-(N-1)A]\dyad{0}+(A+B)\sum_{k=1}^{N-1}\dyad{0}{k}\nonumber\\
     &+(A+B^\ast)\sum_{k=1}^{N-1}\dyad{k}{0}+A\sum_{j,k=1}^{N-1}\dyad{j}{k}\,,
 \end{align}
where
\begin{align}
    A &= \frac{4}{N^2}\sin^2(\omega_N(\lambda)t) \,,\\
    B &= \frac{1}{N}\left(e^{-2it\omega_N(\lambda)}-1\right)\,.
\end{align}

The diagonal elements of $\rho(t)$ provide the probability distribution in Eqs. \eqref{eq:p0_cg}--\eqref{eq:pi_cg}. Instead, the off-diagonal elements allow us to compute the coherence according to Eq. \eqref{eq:coherence_def}, and it reads as
\begin{equation}
    \mathcal{C}(t) = 2 (N-1) \vert A+ B \vert + (N-1)(N-2) \vert A \vert \,.
\end{equation}

\subsection{Star graph}
\label{subapp:an_res_sg}
The star graph has three energy levels (see Table \ref{tab:eig_pbm_star}), thus the unitary time-evolution operator has the following spectral decomposition:
\begin{align}
    \mathcal{U}_\lambda(t) = &\vert  e_0 \rangle \langle e_0 \vert + e^{-2it\omega_{1}} \sum_{l=1}^{N-2} \vert e^l_1 \rangle\langle e^l_1 \vert \nonumber\\
    & + e^{-2it\omega_{N}} \vert e_2 \rangle \langle e_2 \vert\,,
\end{align}
being $\omega_N(\lambda)$ defined in Eq. \eqref{eq:ang_freq}.  Hence, a localized state $\ket{1}$, i.e. an outer vertex, evolves in time according to
\begin{align}
    \vert 1(t) \rangle =&\frac{1}{N}(1-e^{-i2\omega_N(\lambda)t})\ket{0}\nonumber\\
    &+\left(\frac{1}{N}+\frac{N-2}{N-1}e^{-i2\omega_1(\lambda)t}+\frac{e^{-i2\omega_N(\lambda)t}}{N(N-1)} \right)\ket{1}\nonumber\\
    &+\left(\frac{1}{N}-\frac{e^{-i2\omega_1(\lambda)t}}{N-1} +\frac{e^{-i2\omega_N(\lambda)t}}{N(N-1)} \right)\sum_{k=2}^{N-1} \ket{k} \,.
    \label{eq:1t_sg}
\end{align}
Instead, if the initial state is the central vertex $\ket{0}$,
we recover the time evolution of a localized state in the complete graph of the same size (see Eq. \eqref{eq:jt_cg}), and so the same results. Then, the density matrix $\rho(t) = \vert 1(t) \rangle \langle 1(t) \vert$ in the position basis is
\begin{align}
    \rho(t)=&\abs{A}^2\dyad{0}+\abs{B}^2 \dyad{1}+\abs{C}^2 \sum_{k=2}^{N-1}\dyad{k} +\nonumber\\
    &+\Bigg[\abs{C}^2 \sum_{\substack{j,k=2,\\k>j}}^{N-1}\dyad{j}{k}+AB^\ast \dyad{0}{1}\nonumber\\
    &+\sum_{k=2}^{N-1}(AC^\ast\dyad{0}{k}+BC^\ast \dyad{1}{k})+H.c.\Bigg]\,,
    \label{eq:rho_loc_sg}
\end{align}
where $H.c.$ denotes the Hermitian conjugate of the off-diagonal terms only, and
\begin{align}
    A &=\frac{1}{N}(1-e^{-i2\omega_N(\lambda)t})\,,\\
    B &= \frac{1}{N}+\frac{N-2}{N-1}e^{-i2\omega_1(\lambda)t}+\frac{e^{-i2\omega_N(\lambda)t}}{N(N-1)}\,,\\
    C & =\frac{1}{N}-\frac{e^{-i2\omega_1(\lambda)t}}{N-1} +\frac{e^{-i2\omega_N(\lambda)t}}{N(N-1)} \,,
\end{align}
are the coefficients of $\ket{1(t)}$ in the position basis (see Eq. \eqref{eq:1t_sg}). The diagonal elements of $\rho(t)$ provide the probability distribution in Eqs. \eqref{eq:p_10_sg}--\eqref{eq:p_1i_sg}. Instead, the off-diagonal elements allow us to compute the coherence according to Eq. \eqref{eq:coherence_def}. Given the counting of the different matrix elements, and since $\abs{\rho_{j,k}(t)}=\abs{\rho_{k,j}(t)}$, the coherence reads as
\begin{align}
    \mathcal{C} (t) =& 2 \abs{AB^\ast}+ 2(N-2)(\abs{AC^\ast}+\abs{BC^\ast})\nonumber\\
    &+(N-2)(N-3)\abs{C}^2 \,.
\end{align}

It is still pending the issue about the periodicity of the probability distribution. Ultimately, the overall probability distribution is periodic if and only if the periods of the sine functions involved by the probabilities \eqref{eq:p_10_sg}--\eqref{eq:p_1i_sg} are commensurable. Since such sine functions are squared, the periods are:
\begin{align}
    &T_1(\lambda):=\frac{\pi}{\omega_1(\lambda)}=\frac{2\pi}{1+\lambda}\,,\\
    &T_N(\lambda):=\frac{\pi}{\omega_N(\lambda)}=\frac{2\pi}{N+\lambda N^2}\,,\\
    &T_{N,1}(\lambda):=\frac{\pi}{\omega_{N}(\lambda )-\omega_1(\lambda)}=\frac{2\pi}{(N-1)[1+\lambda(N+1)]}\,.
\end{align}
Two non-zero real numbers are commensurable if their ratio is a rational number. The idea is therefore to express both $T_1(\lambda)$ and $T_{N,1}(\lambda)$ as multiple integers of $T_N(\lambda)$. From the ratio  $T_1(\lambda)/T_N(\lambda)$ we get
\begin{equation}
    T_1(\lambda)=\frac{N(1+\lambda N)}{1+\lambda}T_N(\lambda)=:p_N^\lambda T_N(\lambda)\,,
    \label{eq:t1_over_tn_sg}
\end{equation}
with $\lambda \neq -1 \wedge \lambda \neq -1/N$, and from  $T_{N,1}(\lambda)/T_N(\lambda)$ 
\begin{equation}
    T_{N,1}(\lambda)=\frac{N(1+\lambda N)}{(N-1)[1+\lambda(N+1)]}T_N(\lambda)=:q_N^\lambda T_N(\lambda)\,,
    \label{eq:tn1_over_tn_sg}
\end{equation}
with $\lambda \neq -1/N \wedge \lambda \neq -1/(N+1)$. Then, we need to find the value of $\lambda$ such that $p_N^\lambda, q_N^\lambda \in \mathbb{N}$ at the same time. Combining the definition of $p_N^\lambda$ and $q_N^\lambda$ in Eqs. \eqref{eq:t1_over_tn_sg}--\eqref{eq:tn1_over_tn_sg} we find that they are related to $\lambda$ and $N$ by
\begin{equation}
    \lambda = \frac{p_N^\lambda-q_N^\lambda(N-1)}{q_N^\lambda(N^2-1)-p_N^\lambda}\,.
    \label{eq:lambda_sg_pq}
\end{equation}

Please notice that Eq. \eqref{eq:lambda_sg_pq} is to be understood together with Eqs. \eqref{eq:t1_over_tn_sg}--\eqref{eq:tn1_over_tn_sg}. As an example, for $p_N^\lambda=q_N^\lambda$ we get $\lambda=(2-N)/(N^2-2)$ from Eq. \eqref{eq:lambda_sg_pq}. However, the period is unique, so we can not choose any $p_N^\lambda=q_N^\lambda$. Indeed, for such value of $\lambda$ we get $p_N^\lambda=q_N^\lambda=2$ from Eqs. \eqref{eq:t1_over_tn_sg}--\eqref{eq:tn1_over_tn_sg}. In the end, by considering the least common multiple of the latter two integers, the total period of the probability distribution is 
\begin{equation}
    T=\operatorname{lcm}(p_N^\lambda, q_N^\lambda) T_N(\lambda)\,.
    \label{eq:period_sg}
\end{equation}

The above ratios \eqref{eq:t1_over_tn_sg}--\eqref{eq:tn1_over_tn_sg} between the different periods are properly defined unless $\lambda=-1,-1/N,-1/(N+1)$. Nevertheless, for such values of $\lambda$ the overall probability distribution is actually periodic. If we let $p,q\in\mathbb{Z}$, we recover them from Eq. \eqref{eq:lambda_sg_pq} for $q=0$, $q=-p$, and $p=0$, respectively. These values of $\lambda$ make $\omega_1$, $\omega_N$, and $\omega_N-\omega_1$ to vanish, respectively. When $\omega_1=0$ ($\omega_N=0$), the probabilities only involve sine functions with $\omega_N$ ($\omega_1$). When $\omega_N=\omega_1$, all the sine functions have the same angular frequency.

\section{Fisher Information and Quantum Fisher Information for localized states and states maximizing the QFI}
\label{app:q_fi_scg}

In this appendix we prove the analytical results about the Quantum Fisher Information (QFI) in Eq. \eqref{eq:qfi_def_braket} and the Fisher Information (FI) in Eq. \eqref{eq:fi_def} in the different graphs. We provide the FI for a local position measurement whose POVM is given by $\{\dyad{0}, \dyad{1}, \dots, \dyad{N-1}\}$, i.e. by the projectors on the vertex states. The Parthasarathy's Lemma \ref{lemma:parthasarathy} leads to the QFI in Eq. \eqref{eq:qfi_delta_eigenval}, because the state maximizing the QFI involves the ground state and the highest energy eigenstate. The highest energy level might be degenerate, but choosing any eigenstate of such level results in the same QFI. Instead, the FI does depend on such choice.

\subsection{Cycle graph}
\label{subapp:q_fi_cyc}
\paragraph*{Localized state} 
The QFI in Eq. \eqref{eq:qfi_quadratic_time} requires the expectation values of $L^2$ \eqref{eq:L2_cyc} and of
\begin{align}
L^4 =&70{I}+ \sideset{}{'}\sum_{k=0}^{N-1} ( \dyad{k-4}{k}-8\dyad{k-3}{k}\nonumber\\
&+28\dyad{k-2}{k}-56\dyad{k-1}{k}+H.c.)\,,
\end{align}
on the initial state $\ket{j}$. These are $\langle L^2 \rangle=6$ and $\langle L^4 \rangle=70$, from which Eq. \eqref{eq:qfi_loc_cyc} follows.

\paragraph*{States maximizing the QFI} 
In the cycle graph the ground state is unique, whereas the degeneracy of the highest energy level depends on the parity of $N$ (see Table \ref{tab:eig_pbm_cycle}). We define $E_\lambda:=\varepsilon_{max}+\lambda \varepsilon_{max}^2$, where $\varepsilon_{max}$ is the highest energy eigenvalue of $L$ (see Eq. \eqref{eq:eval_cyc_max_even} and Eq. \eqref{eq:eval_cyc_max_odd} for even and odd $N$, respectively).

For even $N$, according to Eq. \eqref{eq:psi0_Nevenodd} the state maximizing the QFI is
\begin{equation}
    \ket{\psi_0(t)} = \frac{1}{\sqrt{2N}}\sum_{k=0}^{N-1}\left[1+(-1)^k e^{-i E_\lambda t} \right]\ket{k}\,,
\end{equation}
and the maximum QFI \eqref{eq:qfi_cycle_evenodd} follows from Eq. \eqref{eq:qfi_delta_eigenval}. Then, the probability distribution associated to a position measurement is
\begin{align}
    P_M(k, t \vert \lambda) = \frac{1}{N}\left[1+(-1)^k \cos(E_\lambda t)\right]\,.
\end{align}
Hence, observing that the dependence on the vertex is encoded only into an alternating sign, the FI is 
\begin{align}
    &\mathcal{F}_c(N,t,\lambda)  =\nonumber\\
    &=\sum_{k=0}^{N/2-1}\left[ \frac{(\partial_\lambda P_M(2k,t\vert \lambda))^2}{P_M(2k,t\vert \lambda)}+\frac{(\partial_\lambda P_M(2k+1,t\vert \lambda))^2}{P_M(2k+1,t\vert \lambda)} \right] \nonumber \\
    &=\frac{\varepsilon_{max}^4 t^2 \sin^2(E_\lambda t)}{N^2}\sum_{k=0}^{N/2-1}\left[ \frac{N}{1+\cos(E_\lambda t)}+\frac{N}{1-\cos(E_\lambda t)}\right]\nonumber\\
    &=\frac{\varepsilon_{max}^4 t^2 \sin^2(E_\lambda t)}{N^2}\frac{N}{2} \frac{2N}{\sin^2(E_\lambda t)}= \varepsilon_{max}^4 t^2 = \mathcal{F}_q(t)\,.
\end{align}
%being $\varepsilon_{max}$ defined in Eq. \eqref{eq:eval_cyc_max_even}.
In other words, the position measurement for the state maximizing the QFI in a cycle graph having an even number of vertices is optimal, since the corresponding FI equals the QFI.

For odd $N$, the situation is trickier: the state maximizing the QFI is not unique, because of the degeneracy of the highest energy level. We may consider the two corresponding eigenstates according to Table \ref{tab:eig_pbm_cycle}, which lead to the following states maximizing the QFI
\begin{equation}
    \ket{\varphi_0^\pm(t)} = \frac{1}{\sqrt{2N}}\sum_{k=0}^{N-1}\left[1+(-1)^k e^{\pm i \theta_k}e^{-i E_\lambda t} \right]\ket{k}\,,
    \label{eq:psi0_cyc_odd_pf}
\end{equation}
 where $\theta_k=\pi k / N$. On the other hand, we may also consider the linear combinations of such eigenstates (see Eqs. \eqref{eq:evec_cyc_max_cos}--\eqref{eq:evec_cyc_max_sin}), which lead to the following states maximizing the QFI
\begin{align}
    \ket{\psi_0^+(t)} &= \frac{1}{\sqrt{2N}}\sum_{k=0}^{N-1}\left[1+\sqrt{2}(-1)^k \cos\theta_k e^{-i E_\lambda t} \right]\ket{k}\label{eq:psi0_cyc_odd_cos}\,,\\
    \ket{\psi_0^-(t)} &=\frac{1}{\sqrt{2N}}\sum_{k=0}^{N-1}\left[1+\sqrt{2}(-1)^k \sin\theta_k e^{-i E_\lambda t} \right]\ket{k}\label{eq:psi0_cyc_odd_sin}\,.
\end{align}
Under the assumption of odd $N$, and according to the following results
\begin{align}
    &\sum_{k=0}^{N-1}\cos^2\theta_k=\sum_{k=0}^{N-1}\sin^2\theta_k=\frac{N}{2}\,,\\
   & \sum_{k=0}^{N-1}(-1)^k e^{\pm i \theta_k}=\frac{1+(-1)^N}{1+e^{\pm i \frac{\pi}{N}}}\stackrel{odd\, N}{=}0\,,
\end{align}
the maximum QFI \eqref{eq:qfi_cycle_evenodd} follows from Eq. \eqref{eq:qfi_delta_eigenval} and does not depend on the choice of these states. Instead, we prove that the FI does depend on them. Again, there is an alternating sign which depends on the vertex. In the following, we will split the sum over even and odd indices, and for odd $N$ it reads as follows:
\begin{equation}
    \sum_{k=0}^{N-1}a_k=\sum_{k=0}^{(N-1)/2}a_{2k}+\sum_{k=0}^{(N-1)/2-1}a_{2k+1}\,.
\end{equation}

We first consider the states $\ket{\varphi_0^\pm (t)}$ in Eq. \eqref{eq:psi0_cyc_odd_pf}. The probability distribution associated to a position measurement is
\begin{equation}
    P_M^\pm(k, t \vert \lambda) = \frac{1}{N}\left[1+(-1)^k \cos(E_\lambda t\mp \theta_k)\right]\,.
\end{equation}
Hence the FI is
\begin{align}
    \mathcal{F}_c^\pm&(\ket{\varphi_0^\pm};N,t,\lambda)=\nonumber\\
    &=\frac{\varepsilon_{max}^4 t^2}{N}\left[\sum_{k=0}^{(N-1)/2} \frac{\sin^2(E_\lambda t\mp \theta_{2k})}{1+\cos(E_\lambda t\mp \theta_{2k})}\right.\nonumber\\
    &\quad\left.+\sum_{k=0}^{(N-1)/2-1} \frac{\sin^2(E_\lambda t\mp \theta_{2k+1})}{1-\cos(E_\lambda t\mp \theta_{2k+1})} \right]\nonumber\\
    & = \frac{\varepsilon_{max}^4 t^2}{N}\left[\sum_{k=0}^{(N-1)/2} (1-\cos(E_\lambda t\mp \theta_{2k}))\right.\nonumber\\
    &\quad\left.+\sum_{k=0}^{(N-1)/2-1} (1+\cos(E_\lambda t\mp \theta_{2k+1})) \right]\nonumber\\
    &= \frac{\varepsilon_{max}^4 t^2}{N}\left[\frac{N-1}{2}+1+\frac{N-1}{2}-1+1 \right]\nonumber\\
    &=\varepsilon_{max}^4 t^2=\mathcal{F}_q(t)\,.
\end{align}
Indeed, for odd $N$
\begin{align}
    &\sum_{k=0}^{(N-1)/2} \cos(\theta_{2k}+\phi)-\sum_{k=0}^{(N-1)/2-1} \cos(\theta_{2k+1}+\phi)\nonumber\\
    &=\sum_{k=0}^{N-1} (-1)^k\cos(\theta_{k}+\phi)=0\,.
\end{align}

Now, we focus on the state $\ket{\psi_0^+(t)}$ in Eq. \eqref{eq:psi0_cyc_odd_cos}. The probability distribution associated to a position measurement is
\begin{align}
    P_M^+(k, t \vert \lambda) &= \frac{1}{2N}\left[1+2\sqrt{2}(-1)^k \cos\theta_k \cos(E_\lambda t)\right.\nonumber\\
    &\quad\left.\vphantom{\sqrt{2}(-1)^k}+2 \cos^2\theta_k )\right]\,.
\end{align}
Hence the FI is
\begin{align}
    &\mathcal{F}_c^+(\ket{\psi_0^+};N,t,\lambda) = \frac{4 \varepsilon_{max}^4 t^2 \sin^2(E_\lambda t)}{N}\nonumber\\
    &\times\left[\sum_{k=0}^{(N-1)/2} \frac{\cos^2\theta_{2k}}{1+2\sqrt{2} \cos\theta_{2k} \cos(E_\lambda t)+2 \cos^2\theta_{2k} }\right.\nonumber\\
    &\quad\left.+\sum_{k=0}^{(N-1)/2-1} \frac{\cos^2\theta_{2k+1}}{1-2\sqrt{2} \cos\theta_{2k+1} \cos(E_\lambda t)+2 \cos^2\theta_{2k+1} }\right]\,.
\end{align}
Analogously for $\ket{\psi_0^-(t)}$ in Eq. \eqref{eq:psi0_cyc_odd_sin}, we find
\begin{align}
    &\mathcal{F}_c^-(\ket{\psi_0^-};N,t,\lambda) = \frac{4 \varepsilon_{max}^4 t^2 \sin^2(E_\lambda t)}{N}\nonumber\\
    &\times\left[\sum_{k=0}^{(N-1)/2} \frac{\sin^2\theta_{2k}}{1+2\sqrt{2} \sin\theta_{2k} \cos(E_\lambda t)+2 \sin^2\theta_{2k} }\right.\nonumber\\
    &\quad\left.+\sum_{k=0}^{(N-1)/2-1} \frac{\sin^2\theta_{2k+1}}{1-2\sqrt{2} \sin\theta_{2k+1} \cos(E_\lambda t)+2 \sin^2\theta_{2k+1} }\right]\,.
\end{align}
Numerical results suggest that $\mathcal{F}_c^\pm(\ket{\psi_0^\pm};N,t,\lambda)<\mathcal{F}_q(t)$. Notice that $\mathcal{F}_c^\pm(\ket{\psi_0^\pm};N,t,\lambda=-1/\varepsilon_{max}) = 0 \,\forall\, t$. Indeed, for such value of $\lambda$ we have that $E_\lambda=0$.

\subsection{Complete graph}
\label{subapp:q_fi_cg}
\paragraph*{Localized state}
The QFI in Eq. \eqref{eq:qfi_quadratic_time} requires the expectation values of $L^2$ and of $L^4$ on the initial state $\ket{0}$. Because of Eqs. \eqref{eq:H0_complete_matrix} and \eqref{eq:Ln_proptoL}, we only need $\langle L \rangle = N-1$, from which Eq. \eqref{eq:qfi_loc_cg} follows.

\paragraph*{States maximizing the QFI} The complete graph has two energy levels: the ground state is unique, but the highest energy level is $(N-1)$-degenerate (see Table \ref{tab:eig_pbm_complete}). The QFI does not depend on the choice of the eigenstate of the highest energy level, but the FI does. As an example, we consider two different states maximizing the QFI $\vert \psi_0^1\rangle$ and $\vert \psi_0^{N-1} \rangle$, i.e. the states in  Eq. \eqref{eq:max_qfi_states_cg} for $l=1$ and $l=N-1$, respectively. 

The first state is
\begin{equation}
 	\vert \psi^0_1(t)\rangle = \frac{1}{\sqrt{2}} \left[\vert e_0 \rangle + \frac{1}{\sqrt{2}}e^{-2it\omega_N(\lambda)}\left(\vert 0 \rangle - \vert 1 \rangle\right)\right]\,.
\end{equation} 
The probability distribution associated to a position measurement is 
\begin{align}
	P_M^0(0,t\vert \lambda) &= \frac{1}{4} + \frac{1}{2N} + \frac{\cos\left(2t \omega_N(\lambda)\right)}{\sqrt{2N}}\,, \\
	P_M^0(1,t\vert \lambda) &= \frac{1}{4} + \frac{1}{2N} - \frac{\cos\left(2t \omega_N(\lambda)\right)}{\sqrt{2N}}\,, \\
	P_M^0(k,t\vert \lambda) &= \frac{1}{2N}\,, \text{with $2\leq k \leq N-1$.}
\end{align}
Then, being null the $(N-2)$ contributions from the vertices $2\leq k \leq N-1$, since $\partial_\lambda (P_M^0(k,t\vert \lambda))= 0$, only the probabilities associated to the vertices $\ket{0}$ and $\ket{1}$ contribute to the FI \eqref{eq:fi_def}, which results in Eq. \eqref{eq:fi_max_qfi_states_cg_1}.

Similarly, the second state is
\begin{align}
	\vert \psi^{N-1}_0 (t) \rangle  = &\frac{1}{\sqrt{2}} \left\lbrace\vert e_0 \rangle + \frac{1}{\sqrt{N^2-N}}e^{-2it\omega_N(\lambda)}\left[\vert 0 \rangle+\dots\right.\right. \nonumber \\
	&  \left.\vphantom{\frac{}{}}\left.+ \vert N-2 \rangle - (N-1) \vert N-1 \rangle \right]\right\rbrace\,.
\end{align}
The probability distribution associated to a position measurement is
\begin{equation}
	P^{N-1}_M (k,t\vert \lambda) =\frac{1}{2(N-1)}  + \frac{1}{N\sqrt{N-1}}\cos\left(2t\omega_N(\lambda)\right)
	\end{equation}
with $0\leq k\leq N-2$, and
\begin{equation}
	P^{N-1}_M(N-1,t\vert \lambda) = \frac{1}{2}-\frac{\sqrt{N-1}}{N}\cos(2t\omega_N(\lambda))\,.
\end{equation}
Then, having $(N-1)$ equal contributions from the vertices $0\leq k \leq N-2$ and a particular one from $N-1$, the FI \eqref{eq:fi_def} results in Eq. \eqref{eq:fi_max_qfi_states_cg_2}.

\subsection{Star graph}
\label{subapp:q_fi_sg}
\paragraph*{Localized state} Considering the central vertex $\ket{0}$ as the initial state provides the same results observed in the complete graph of the same size. Thus, we consider as initial state $\ket{1}$, i.e. one of the outer vertices The QFI in Eq. \eqref{eq:qfi_quadratic_time} requires the expectation values of $L^2$ \eqref{eq:L2_star} and of
\begin{align}
L^4=&(N^2+N+2)I-N^3 \sum_{k=1}^{N-1}(\dyad{k}{0}+\dyad{0}{k})\nonumber\\
&+(N-2)(N^3+N^2+N+1)\dyad{0}\nonumber\\
&+(N^2+N+1)\sum_{\substack{j,k=1,\\j\neq k}}^{N-1}\dyad{j}{k}
\end{align}
on the initial state $\ket{1}$. These are $\langle L^2 \rangle=2$ and $\langle L^4 \rangle=N^2+N+2$, from which Eq. \eqref{eq:qfi_loc_sg} follows. 

\paragraph*{States maximizing the QFI} In the star graph the state maximizing the QFI, according to Eq. \eqref{eq:max_qfi_states_sg}, is
\begin{equation}
	\vert \psi_0 (t) \rangle = \frac{1}{\sqrt{2}} \left(\vert e_0 \rangle + e^{-2it\omega_N(\lambda)}\vert e_2 \rangle\right)\,,
\end{equation}
being both the ground and the highest energy levels not degenerate (see Table \ref{tab:eig_pbm_star}). Then, the probability distribution associated to a position measurement is
\begin{align}
    P_M(0, t \vert \lambda) &= \frac{1}{2} + \frac{\sqrt{N-1}}{N} \cos(2 t \omega_{N}(\lambda))\,, \\
    P_M(k, t\vert \lambda) &= \frac{1}{2(N-1)} - \frac{1}{N\sqrt{N-1}}\cos(2t\omega_N(\lambda))\,,
\end{align}
with $1 \leq k \leq N-1$. Then, the FI follows from Eq. \eqref{eq:fi_def}.

\subsection{Maximum QFI states: the role of the phase factor in the superposition of energy eigenstates}
\label{subapp:phi_maxQFI_states}
So far we have studied the states maximizing the QFI without bothering to consider a different linear combination of the ground state and the highest energy state. According to the Parthasarathy's lemma \ref{lemma:parthasarathy}, the two eigenstates defining the state in Eq. \eqref{eq:psi_0} are equally weighted. However, we may suppose the second one to have a phase factor, i.e
\begin{equation}
    	\vert \psi_0 \rangle = \frac{1}{\sqrt{2}}(\vert e_0 \rangle + e^{i\phi} \vert e_1 \rangle)\,.
\end{equation}
In this section, we study how the phase $\phi$ affects the FI and QFI.

The states $\ket{e_0}$ and $\ket{e_1}$ denote the eigenstates of minimum and maximum energy eigenvalue, i.e. $\varepsilon_{min}$ and $\varepsilon_{max}$ respectively, and we know that for simple graphs $\ket{e_0}=(1,\ldots,1)/\sqrt{N}$ and $\varepsilon_{min}=0$. Moreover, since the Laplacian matrix is real and symmetric, we can always deal with real eigenstates. Because of Eq. \eqref{eq:qfi_quadratic_time}, we already know that the QFI is \eqref{eq:qfi_delta_eigenval} and therefore it is independent of a phase shift. On the other hand, the FI reads as follows:
\begin{align}
	&\mathcal{F}_c(t, \lambda) = 2 t^2 \varepsilon_{max}^4 \sin^2(E_\lambda t - \phi)\nonumber\\
	&\times \sum_{i=0}^{N-1}\frac{\langle i \vert e_1 \rangle^2}{N\langle i \vert e_1 \rangle^2 +  2\sqrt{N}\langle i \vert e_1 \rangle \cos(E_\lambda t - \phi) +1}\,,
\end{align}
where $E_\lambda:=\varepsilon_{max}+\lambda \varepsilon_{max}^2$, and $\langle i \vert e_1 \rangle\in \mathbb{R}$, since the vectors involved are real. Hence, the phase is encoded as a phase shift in all the sine and cosine functions. However, this does not result in a global time shift, because the quadratic term in $t$ is not affected by $\phi$.

\bibliography{qwp4_bib}

\end{document}